\documentclass[10pt,pra,floatfix,twocolumn,amsfonts,amsmath,amssymb,superscriptaddress,nofootinbib]{revtex4-1}

\usepackage[colorlinks=true,citecolor=blue,urlcolor=blue]{hyperref}
\usepackage[utf8]{inputenc}
\usepackage[T1]{fontenc}
\usepackage[sc,osf]{mathpazo}
\usepackage{amsmath, amsthm, amssymb,amsfonts}
\usepackage{graphicx}
\usepackage{dcolumn}
\usepackage{bm}
\usepackage{bbm}
\usepackage{mathtools}
\usepackage{comment}
\usepackage{color}
\usepackage{xcolor}

\usepackage{graphicx}
\usepackage{amssymb}
\usepackage{amsmath}
\usepackage{amsthm}
\usepackage{psfrag}
\usepackage{epsfig}
\usepackage{bbm}
\usepackage{bm}
\usepackage{dsfont}
\usepackage[normalem]{ulem}
\hypersetup{breaklinks}

\newcommand{\beq}{\begin{eqnarray}}
\newcommand{\eeq}{\end{eqnarray}}

\newcommand{\CI}{{\perp\!\!\!\perp}}

\newtheorem*{theorem*}{Theorem}

\DeclareMathOperator{\conv}{conv}

\definecolor{green}{RGB}{7,133,10}

\begin{document}
\title{The Entropic Approach to Causal Correlations}

\author{Nikolai Miklin}
\thanks{These authors contributed equally to this work.}
\affiliation{Naturwissenschaftlich-Technische Fakult\"{a}t, Universit\"{a}t Siegen, Walter-Flex-Strasse 3, 57068 Siegen, Germany}

\author{Alastair A. Abbott}
\thanks{These authors contributed equally to this work.}
\affiliation{Institut N\'{e}el, CNRS and Universit\'{e} Grenoble Alpes, 38042 Grenoble Cedex 9, France}

\author{Cyril Branciard}
\affiliation{Institut N\'{e}el, CNRS and Universit\'{e} Grenoble Alpes, 38042 Grenoble Cedex 9, France}

\author{Rafael Chaves}
\affiliation{International Institute of Physics, Federal University of Rio Grande do Norte, 59070-405 Natal, Brazil}

\author{Costantino Budroni}
\affiliation{Institute for Quantum Optics and Quantum Information (IQOQI), Boltzmanngasse 3 1090 Vienna, Austria}

\date{September 19, 2017}

\begin{abstract}
	The existence of a global causal order between events places constraints on the correlations that parties may share.
	Such ``causal correlations'' have been the focus of recent attention, driven by the realization that some extensions of quantum mechanics may violate so-called causal inequalities.
	In this paper we study causal correlations from an entropic perspective, and we show how to use this framework to derive entropic causal inequalities. 
	We consider two different ways to derive such inequalities. 
	Firstly, we consider a method based on the causal Bayesian networks describing the causal relations between the parties. 
	In contrast to the Bell-nonlocality scenario, where this method has previously been shown to be ineffective, we show that it leads to several interesting entropic causal inequalities. 
	Secondly, we consider an alternative method based on counterfactual variables that has previously been used to derive entropic Bell inequalities. 
	We compare the inequalities obtained via these two methods and discuss their violation by noncausal correlations. 
	As an application of our approach, we derive bounds on the quantity of information -- which is more naturally expressed in the entropic framework -- that parties can communicate when operating in a definite causal order.
\end{abstract}

\maketitle

\section{Introduction}

When describing most physical phenomena it seems natural to assume that physical events take place in a well-defined causal structure. 
For instance, earlier events can influence later ones but not the opposite, or, if two events are distant enough (typically, space-like separated) from each other, any correlation between them can only be due to some common cause in their past. This intuition is formalized in Reichenbach's principle~\cite{Reichenbachbook} and generalized by the mathematical theory of causal models~\cite{Pearlbook} that form the basis for our current understanding of how to infer causation from empirically observed correlations. 
Not surprisingly, it has found a wide range of applications~\cite{Pearlbook,Angrist1996,Friedman2004}. 
Yet, quantum phenomena defy such an intuitive notion of cause and effect.

As shown by Bell's Theorem~\cite{Bell1964}, quantum correlations obtained by measurements on distant entangled parties are incompatible with Reichenbach's principle~\cite{Cavalcanti2014,Wood2015} or, more generally, with classical theories of causality, forcing us to generalize the notion of causal models~\cite{Leifer2013,Henson2014,Chaves2015,Ried2015,Costa2016,Allen2016}. 
In a scenario where different experimenters interact only once with a given system that is exchanged between them, one could expect that no simultaneous causal influences between them should be possible but rather only one-way influences. 
However, it has been realized that physical theories do not necessarily have to comply with the idea of a definite causal order~\cite{oreshkov12,Portmann2017}. 
One can also imagine theories where the causal order itself is in a sort of ``quantum superposition''~\cite{oreshkov12,Chiribella2013}, which can be verified using so-called causal witnesses~\cite{Araujo2015,Branciard2016}. 

As for entanglement witnesses~\cite{horodecki96,Horodecki2009}, the use of causal witnesses assumes that we have a precise description of the measurement apparatus, that is, they are relevant in a device-dependent framework. Nevertheless, by allowing physical theories that are locally equivalent to quantum mechanics but relax the assumption of a fixed global causal structure, it is possible to verify causal indefiniteness also in a device-independent manner. With the aim of providing a general framework to such scenarios, the process matrix formalism~\cite{oreshkov12} has been introduced and shown to allow for the violation of so-called causal inequalities~\cite{oreshkov12,baumeler13,baumeler14,oreshkov15,Branciard:2016aa,Abbott:2016aa}, which are device-independent constraints that play a similar role to that of Bell inequalities~\cite{Bell1964}. However, whether violations of causal inequalities can be experimentally observed is still an important open question.

Our goal in this paper is to introduce a new framework for the derivation of causal inequalities and the study of their potential violations: the entropic approach to causal correlations. The idea of using entropies to understand sets of correlations has its origins in the context of Bell inequalities~\cite{Braunstein1988,Chaves:2012aa,Fritz:2013aa,Chaves2014} but since then has also found various other applications in quantum contextuality~\cite{Kurzynski2012,Chaves:2013aa,Raesi2015}, device-independent applications~\cite{Chaves2015DI,Zhu2016}, causal inference~\cite{Chaves2014b,Henson2014,Pienaar2016} and in the characterization of nonsignaling correlations~\cite{Chaves:2016aa}. As for these previous applications, the interest in characterizing the entropies compatible with causal correlations stems not only from practical and technical issues, but also from a more fundamental point. 
To begin with, causal inequalities expressed in terms of probabilities are constructed for a fixed number of inputs and outputs, and their systematic derivation becomes harder as this number increases~\cite{Branciard:2016aa,Abbott:2016aa}. 
In contrast, we will derive entropic causal inequalities that are valid for arbitrary finite alphabets either for the input and output variables, or just for the output variables. Furthermore, entropic inequalities can be easily combined with extra assumptions, such as conditional independence relations or information theoretic constraints (e.g., bounds on the amount of communication), which would be hard to treat in the probabilistic framework~\cite{Chaves2014b,Chaves:2016aa,Chaves2016Po}. 
More fundamentally, given that entropies are a core concept in classical and quantum information theory, it is of clear relevance to have a framework that focuses on these quantities rather than on probabilities, and it may help connect causal inequalities with principles such as information causality~\cite{Pawlowski2009}.

The paper is organized as follows. 
In Sec.~\ref{sec:preliminary}, we will introduce the basic notions relevant for our investigation, namely causal correlations and the entropic approach to causal structures, and elaborate two complementary ways in which the approach can be applied.
In Sec.~\ref{sec:bipartite} we will show how to derive entropic causal inequalities for the bipartite scenario, and discuss 
their violation. 
In Sec.~\ref{sec:multipartite}, we will explain how this approach can be generalized to multipartite scenarios.
Finally, as an application, in Sec.~\ref{sec:infoBounds} we use this approach to derive bounds on mutual informations in causal games.

\section{Preliminaries}\label{sec:preliminary}

\subsection{Causal correlations}
\label{sec:causalcorr}

Causal correlations are most easily introduced in the bipartite case, where we consider two parties, Alice (${\rm A}$) and Bob (${\rm B}$), who together conduct a joint experiment while each having control over a separate closed laboratory.
During each round of the experiment, Alice and Bob each receive, operate on, and send out a single physical system, which is the only means by which they may communicate.
In addition, they each receive some (external) classical inputs $X$ and $Y$, for Alice and Bob respectively, and produce some classical outputs $A$ and $B$, respectively. 
Throughout the paper we use upper-case letters (e.g., $X$) to denote random variables, and corresponding lower-case letters (e.g., $x$) to denote the specific values they take.
Their probability distributions will generically be denoted by $P$; we will also use the shorthand notations $P(x)$ for $P(X=x)$, $P(x_{\text{\tiny (}},_{\text{\tiny )\!\!}}y)$ for $P(X=x,Y=y)$, $P(a|x)$ for $P(A=a|X=x)$, etc.

The joint conditional probability distributions $P(ab|xy)$ that can be produced in such an experiment depend on the causal relation between Alice and Bob.
If Bob cannot signal to Alice their correlations should obey $P(a|xy)=P(a|xy')$ for all $x,y,y',a$, where $P(a|xy)=\sum_bP(ab|xy)$.
We denote this situation by $\mathrm{A}\prec\mathrm{B}$, and write $P = P^{\mathrm{A}\prec \mathrm{B}}$ in this case. 
Note that this does not necessarily imply that Alice is in the causal past of Bob since the events could be space-like separated, but merely that the correlation is compatible with such a causal order.
Similarly, if the correlation is compatible with Bob being in the causal past of Alice we write $\mathrm{B}\prec \mathrm{A}$ and we have $P^{\mathrm{B}\prec \mathrm{A}}(b|xy)=P^{\mathrm{B}\prec \mathrm{A}}(b|x'y)$ for all $x,x',y,b$.
The correlations that satisfy both these conditions (and are thus consistent both with $\mathrm{A}\prec \mathrm{B}$ and $\mathrm{B}\prec \mathrm{A}$) are precisely the nonsignaling correlations~\cite{Popescu1994}.

More generally, we are interested in the correlations achievable under the assumption of a definite causal order in  each round of the experiment, even if the causal relation between Alice and Bob may be different (e.g., chosen randomly) for each individual round. We thus say that a correlation $P(ab|xy)$ is \emph{causal} if it can be written as
\begin{equation}\label{eq:causalCorreltn}
	P(ab|xy)=q_0\,P^{\mathrm{A}\prec \mathrm{B}}(ab|xy) + q_1\,P^{\mathrm{B}\prec \mathrm{A}}(ab|xy),
\end{equation}
with $q_0,q_1\in [0,1]$ and $q_0+q_1=1$, where $P^{\mathrm{A}\prec \mathrm{B}}(ab|xy)$ and $P^{\mathrm{B}\prec \mathrm{A}}(ab|xy)$ satisfy the respective (one-way) no-signaling conditions defined above~\cite{oreshkov12}.  

It was shown in Ref.~\cite{Branciard:2016aa} that the set of bipartite causal correlations forms a convex polytope, whose vertices are simply the deterministic causal correlations (i.e., causal correlations for which the outputs $A,B$ are deterministic functions of the inputs $X,Y$).
The facets of this polytope specify \emph{causal inequalities}, analogous to Bell inequalities for local correlations, that any causal correlation must satisfy~\cite{oreshkov12}.
The situation with binary input and output variables was characterized completely in~\cite{Branciard:2016aa}, where it was shown that there are only two nonequivalent causal inequalities (up to symmetries).
The simplest of these is perhaps the ``guess your neighbor's input'' (GYNI) inequality, which has a simple interpretation as a game (up to a relabeling of the inputs and outputs) in which  the inputs $X,Y$ are chosen uniformly at random and the goal is for each party to output the other party's input.
One such form of this inequality can be written~\cite{Branciard:2016aa}
\begin{equation}\label{eq:GYNI}
	\frac{1}{4}\sum_{x,y,a,b}\delta_{a,y}\,\delta_{b,x}\,P(ab|xy)\le \frac{1}{2},
\end{equation}
where $\delta$ is the Kronecker delta function.

The notion of causal correlations can be generalized to more parties, although one has to take into account the fact that, in a given round of the experiment, the causal order of some parties may depend on the inputs and outputs of previous parties~\cite{oreshkov15,Abbott:2016aa}.
In this paper we will primarily, in Sec.~\ref{sec:bipartite}, focus on applying the entropic approach to bipartite causal correlations, before returning to the multipartite case in Sec.~\ref{sec:multipartite}.

\subsection{The entropic approach and marginal problems}
\label{sec:entropymarginal}

Below we introduce the basic notions concerning entropy cones and marginal scenarios. 
We then review the entropic characterization of marginal scenarios~\cite{Fritz:2013aa} using two complementary methods, the first considering the entropies of the variables composing a given causal model, and the second based on the counterfactual approach to correlations.
To illustrate concretely and contrast these two methods, we apply them to the well-known Bell scenario.
Readers well-familiarized with the entropic approach may prefer to skip these expository examples.

\subsubsection{Entropy and Shannon cones}

Let $S=\{X_1, \dots, X_n\}$ be a set of $n$ random variables taking values $x_1, \dots, x_n$, whose joint distribution $P(x_1, \dots, x_n)$ we wish to characterize entropically.
For every nonempty subset $T\subset S$ we shall denote by $\bm{X}_T = (X_i)_{X_i \in T}$ the joint random variable that involves all variables in $T$, taking values $\bm{x}_T = (x_i)_{X_i \in T}$. 
We can then compute the marginal \emph{Shannon entropies} $H(\bm{X}_T)=H(T)$ from the marginal probability distributions $P(\bm{X}_T=\bm{x}_T)=P(\bm{x}_T)$ as
\begin{equation}
	\label{ShannonE}
	H(T)\coloneqq -\sum_{\bm{x}_T}P(\bm{x}_T)\log_2{P(\bm{x}_T)}.
\end{equation}
Together with $H(\emptyset)\coloneqq 0$, every global probability distribution $P(x_1,\ldots,x_n)$ thus specifies $2^n$ real numbers in the entropic description, which can be expressed as the components of a $(2^n)$-dimensional vector ${\bm{h}=(H(\emptyset),H(X_1),\ldots,H(X_{1}X_{2}),\ldots, H(X_1 \dots X_n))} = (H(T))_{T \subset S}$ in $\mathds{R}^{2^n}$.

A fundamental problem in information theory is to decide whether a given vector is an \emph{entropy vector}, that is, if it is obtainable from some probability distribution. 
The (closure of the) region of valid entropy vectors
\begin{equation}\label{eq:entropyCone}
\Gamma_{S}^* \coloneqq	\overline{\left\{ \bm{h} \in \mathds{R}^{2^n} \,|\, \bm{h} = (H(T))_{T\subset S} \right\}},
\end{equation}
is known to be a convex cone, called the \emph{entropy cone} (see Ref.~\cite{Yeung2008} for a comprehensive discussion of entropy cones). 
There is no known explicit description of $\Gamma_{S}^*$, so one generally has to rely on an approximation of it. 
A well-known and very useful outer approximation of $\Gamma_{S}^*$ is the so-called \emph{Shannon cone} $\Gamma_{S}$, defined by the elemental inequalities
\begin{align}
\label{shannonineqs_basic}
	H(S\setminus\{X_i\}) &\leq H(S), \\ 
	H(T) + H(T\cup\{X_i,X_j\}) &\leq H(T\cup \{X_i\}) + H(T\cup \{X_j\}),\notag
\end{align}
for all $1\le i, j\le n$, $i \neq j$, and $T \subset S \setminus\{X_i,X_j\}$. 
That is, the Shannon cone  $\Gamma_{S}$ is described by a finite system of $m=n+2^{n-2}\binom{n}{2}$ linear inequalities, which one can write in the form $I\bm{h}\leq \bm{0}$, where $I$ is an $m\times 2^n$ real matrix and $\bm{0}$ a vector with null entries.
The inequalities in Eq.~\eqref{shannonineqs_basic} are the minimal set of inequalities implying the monotonicity of entropy, i.e., $H(U|T) \coloneqq H(TU) - H(T) \geq 0$, and the submodularity (or strong subadditivity), i.e., $I(U:V|T)\coloneqq H(TU)+H(TV)-H(TUV)-H(T)\geq 0$, for any subsets $T,U,V\subset S$.  
These inequalities and any combination thereof are known as \emph{Shannon-type inequalities}.
It is known that for $n\le 3$ variables every inequality delimiting the entropy cone $\Gamma_{S}^*$ is of the Shannon type; however, this is not the case for $n> 3$~\cite{Yeung2008}.

The inequalities characterizing the Shannon cone simply arise from demanding that the function $P(\bm{x}_T)$  appearing in~\eqref{ShannonE} should be identified with a valid probability distribution (i.e., it should be nonnegative and normalized). 
However, one often wishes to consider (and characterize the entropy vectors for) situations where additional constraints on the random variables are known.
For example, $X_i$ and $X_j$ might be known to be independent, which implies that $P(x_i,x_j)=P(x_i)P(x_j)$. 
Such independence constraints, which are nonlinear in terms of probabilities, define simple linear constraints in terms of entropies, e.g., $P(x_i,x_j)=P(x_i)P(x_j) \rightarrow H(X_iX_j)=H(X_i)+H(X_j)$. 
These extra constraints can be easily incorporated into the entropic framework since they define a linear subspace, which we denote $\textrm{L}_{\mathcal{C}}$, characterized by linear equalities.
When combined with the elemental inequalities one obtains a new finite system of inequalities $I^{\prime}\bm{h}\leq \bm{0}$ characterizing the ``constrained Shannon cone'' $\Gamma_{S}\bigcap \textrm{L}_{\mathcal{C}}$. 

In some cases, one may also wish to add linear inequality constraints which, in general, may give rise to more general polyhedra described by inhomogeneous systems of linear inequalities $I^{\prime}\bm{h}\leq \bm{\beta}$~\cite{Schrijver_book}. 
In such cases we will again denote the set of vectors $\bm{h}$ satisfying these additional constraints as $\textrm{L}_{\mathcal{C}}$; we will return to this point in more detail in Sec.~\ref{sec:bipartite}.

\subsubsection{Marginal scenarios}
\label{subsec:marginal}

Consider again a set of random variables $\{X_1,\dots,X_n\}$ with a joint probability distribution $P(x_1,\ldots,x_n)$. 
We often encounter situations where not all variables, or combinations thereof, are empirically accessible. 
For example, our system of interest could be composed of three random variables $X_1,X_2,X_3$ but, for some reason, we can access at most two of them at a time, thus implying that we cannot know their joint entropy $H(X_1X_2X_3)$. 
Alternatively, there might be variables that represent latent  factors~\cite{Pearlbook} and that, for this reason, are unobservable.
In such cases, we face a \emph{marginal problem}: decide whether some given information on the marginals is compatible with a global description fulfilling certain constraints (for example the elemental entropy inequalities). 
In the example with three variables, it is easy to see that the elemental inequalities imply that
\begin{equation}
H(X_1) + H(X_2X_3) \leq H(X_1X_2)+H(X_1X_3).
\end{equation}
That is, the global structure of entropy vectors implies nontrivial constraints (which are not elemental inequalities~\eqref{shannonineqs_basic}) that should be respected by any marginal information compatible with it.

More formally, given a set of random variables $S=\{X_1,\ldots,X_n\}$, a \emph{marginal scenario} is a collection of subsets $\mathcal{M}=\{ M_1,\ldots,M_{|\mathcal{M}|}\}$, $M_j\subset S$ representing those variables for which we have access to the probability distribution $P(\bm{x}_{M_j})$ (and thus to $H(M_j)$).
Clearly, $M_j\in \mathcal{M}$ and $M_j'\subset M_j$ implies $M_j'\in \mathcal{M}$, that is, given some probability distribution we also have access to any marginal of it.
In a slight abuse of notation we will therefore write $\mathcal{M}$ only in terms of its maximal subsets, since these are sufficient to specify the entire marginal scenario; the complete representation of $\mathcal{M}$, which explicitly includes all (not necessarily maximal) subsets $T$ for which the marginal distribution $P(\bm{x}_{T})$ is accessible, will be denoted $\mathcal{M}^{\rm c} = \{T \mid T\subset M_j, M_j \in \mathcal{M}\}$.
In the example above the marginal scenario would then be represented as $\mathcal{M}=\big\{ \{X_1,X_2\},\{X_1,X_3\},\{X_2,X_3\}\big\}$, or $\mathcal{M}^{\rm c}=\big\{ \emptyset, \{X_1\}, \{X_2\}, \{X_3\}, \{X_1,X_2\},\{X_1,X_3\},\{X_2,X_3\}\big\}$.

In general we are interested in characterizing the entropy cone $\Gamma^*_{\mathcal{M}}$ associated with a marginal scenario $\mathcal{M}$, thus obtaining constraints implied by the global entropy cone on the marginal subspace of interest. 
Geometrically, this corresponds to the projection of the original entropy cone onto the subspace of entropy vectors $\bm{h}=(H(T))_{T \in \mathcal{M}^{\rm c}} \in \mathds{R}^{|\mathcal{M}^{\rm c}|}$, corresponding to the variables in $\mathcal{M}^{\rm c}$. 
Since, in practice, we work with the Shannon cone $\Gamma_{S}$ -- possibly constrained by some further linear constraints specifying a subset of entropy vectors $\textrm{L}_{\mathcal{C}}$, as described previously -- which is characterized by a finite system of inequalities, this projection corresponds to a simple variable elimination of all the terms not contained in $\mathcal{M}^{\rm c}$~\cite{Williams1986,budroni2012bell2,Fritz:2013aa}. 
After removing redundant inequalities, the remaining inequalities are facets (i.e., the boundaries) of the Shannon cone, or more generally polyhedron, in the observable marginal subspace. 
Formally, the marginal Shannon polyhedron $\Gamma_{\mathcal{M}}$ is defined as
\begin{equation}
\label{eq:ProjMargCone}
\Gamma_{\mathcal{M}}=\Pi_\mathcal{M}\left(\Gamma_{S} \bigcap \textrm{L}_{\mathcal{C}} \right),
\end{equation}
where $\Pi_\mathcal{M}$ denotes the projection onto the coordinates associated with the marginal scenario $\mathcal{M}$ -- i.e., onto the coordinates $H(T)$ with $T \in \mathcal{M}^{\rm c}$. 

\subsubsection{Probability structures}
\label{sec:probStructures}

The characterization of entropy cones (or polyhedra) and marginal problems outlined above can be easily extended to the case where we no longer assume that there is a well-defined global probability distribution over all the variables in the set $S$.
Instead, we may assume that only certain subsets of variables have such a joint distribution, and that only the marginals of certain subsets of these subsets are empirically accessible.
This type of restriction may be imposed by assumptions about the underlying physical theory being described, as will be clear in the example we discuss in Sec.~\ref{sec:counterfactuals}.

We will denote the collection of subsets of $S$ for which we assume joint probability distributions exist by $\mathcal{S} = \{S_1,\ldots,S_{|\mathcal{S}|}\}$, with each $S_i \subset S$ such that $\cup_i S_i = S$; we call $\mathcal{S}$ the \emph{probability structure}.\footnote{Of course the probability assignment should be consistent. That is, for two subsets $S_i$ and $S_i'$ of $\mathcal{S}$, the corresponding probability distributions $P_i$ and $P_i'$ should coincide on $S_i \cap S_i'$, so that one must have $P_i(\bm{x}_{T}) = P_i'(\bm{x}_{T})$ for all $T \subset S_i \cap S_i'$. This allows one to define $H(T)$ for all $T \in \mathcal{S}^{\rm c}$ unambiguously.}
As for the marginal scenario, we will represent $\mathcal{S}$ by just its maximal subsets in a slight abuse of notation; the complete representation of $\mathcal{S}$, that explicitly includes all subsets for which a joint probability distribution exists, will similarly be denoted $\mathcal{S}^{\rm c}$.
In such a situation the entropies $H(T)$ cannot be defined for all subsets $T \subset S$, but only for the subsets in $\mathcal{S}^{\rm c}$. The entropy vectors we shall consider will thus be defined here as $\bm{h}=(H(T))_{T \in \mathcal{S}^{\rm c}} \in \mathds{R}^{|\mathcal{S}^{\rm c}|}$.
Again, no explicit characterization is known for the set of valid entropy vectors; we will instead rely on its outer approximation characterized via the Shannon constraints, now restricted to each subset $S_i \in \mathcal{S}$. Namely, the Shannon cone of interest is now
\begin{equation}\label{eq:defconeS}
	\Gamma^{\mathcal{S}}=\bigcap_{S_i\in \mathcal{S}} \Gamma_{S_i},
\end{equation} 
where $\Gamma_{S_i}\subset \mathds{R}^{|\mathcal{S}^{\rm c}|}$ is the cone defined by the Shannon inequalities on the variables in $S_i$, which, in particular, leave the other variables in $S \setminus S_i$ unconstrained. 
In the extremal case where we do assume a global joint probability distribution for all variables we have $\mathcal{S}=\{S\}$, $\mathcal{S}^{\rm c}=2^S$, and we recover $\Gamma^{\mathcal{S}}=\Gamma_{S}$.

One can similarly consider marginal scenarios under a given probability structure $\mathcal{S}$, with the constraint that marginals must arise from existing probability distributions, i.e., for all $M_j\in \mathcal{M}$ there must exist an $S_i\in \mathcal{S}$ such that $M_j\subset S_i$. 
One can also add linear constraints to the entropy vectors under consideration, as before, represented by some subset of entropy vectors $\textrm{L}_{\mathcal{C}}$. 
We can thus define the marginal Shannon polyhedron associated with $\mathcal{S},\mathcal{M}$, and $\textrm{L}_{\mathcal{C}}$ as
\begin{equation}
	\label{eq:ProjMargConeS}
	\Gamma_{\mathcal{M}}^{\mathcal{S}}=\Pi_\mathcal{M}\left(\Gamma^{\mathcal{S}} \bigcap \textrm{L}_{\mathcal{C}} \right).
\end{equation} 

The choice of probability structure can generally be considered on a case-by-case basis depending on the scenario being modeled.
Unless otherwise stated we will take $\mathcal{S}=\{S\}$ but, as we will discuss, this will not always be the most pertinent choice.

\subsubsection{The entropic characterization of causal Bayesian networks}
\label{sec:causal_str}

In order to describe the causal relations between random variables, we will first use the framework of causal Bayesian networks\footnote{Note that although notions of causal correlations and causal Bayesian networks both share the ``causal'' qualifier, they are distinct concepts: a causal correlation is \emph{not} simply one that can be obtained from any particular causal Bayesian network.}~\cite{Pearlbook}.
Such networks can be conveniently represented as \emph{directed acyclic graphs} (DAGs), in which each node represents a variable and directed edges (arrows) encode the causal relations between them. 
A set of variables $S=\{X_1,\dots,X_n\}$ forms a Bayesian network with respect to a given DAG if and only if the variables admit a global probability distribution $P(x_1,\dots,x_n)$, i.e., $\mathcal{S}=\{ S\}$, that factorizes according to
\begin{equation}
P(x_1,\dots,x_n)= \prod_{i=1}^{n} P(x_i\vert \mathrm{Pa}_i),
\end{equation}
where $\mathrm{Pa}_i$ stands for the graph-theoretical parents of variable $X_i$, that is, all those variables $X_j$ that have an outgoing edge pointing to $X_i$ in the DAG under consideration. 
The decomposition above implies a set of conditional independences (CIs), which are either independence relations of the type $P(x_i,x_j)=P(x_i)P(x_j)$ (in which case we write $X_i \CI X_j$) or conditional independence relations such as $P(x_i,x_j | x_k)=P(x_i\vert x_k)P(x_j \vert x_k)$ (denoted $X_i\CI X_j \mid X_k$).\footnote{For CIs between more than two variables, we use the natural extension of this notion. For example, if $P(x_i,x_j,x_k)=P(x_i)P(x_j)P(x_k)$ we write $X_i \CI X_j \CI X_k$.} 
Given a DAG, a complete list of CIs can be obtained via the $d$-separation criterion~\cite{Pearlbook}.
If the arrows in the DAG representation of a Bayesian network describe the direct causal relations between the variables in question, then we call it a \emph{causal Bayesian network}. 

Entropically, these CIs correspond to simple linear relations: $X_i \CI X_j \rightarrow H(X_iX_j)=H(X_i)+H(X_j)$ and $X_i\CI X_k \mid X_k \rightarrow H(X_iX_j\vert X_k)=H(X_i\vert X_k)+H(X_j \vert X_k)$. 
As a result, the set of entropy vectors compatible with a given DAG is the intersection of the entropy cone $\Gamma_{S}^*$ with the linear subspace $\textrm{L}_{\mathrm{CI}}$ defined by the set of linear constraints that characterize the CIs associated with the DAG~\cite{Chaves2014,Chaves2014b}. 
In practice, we again rely on the outer approximation given by the intersection of the Shannon cone $\Gamma_{S}$ with $\textrm{L}_{\mathrm{CI}}$.

If all the variables in a DAG are observable, in order to check the compatibility of a given entropy vector with the DAG it suffices to check whether all the entropic CIs are satisfied. 
However, we are often interested in DAGs containing latent, non-observable, variables. 
Splitting the $n$ variables making up the DAG into $j$ observable variables $O_1,\dots,O_j$ and $n-j$ latent variables $\Lambda_1,\dots,\Lambda_{n-j}$ we thus need to compute the marginal Shannon cone $\Pi_\mathcal{M}\left(\Gamma_{S} \bigcap \textrm{L}_{\mathrm{CI}} \right)$ where $\mathcal{M}=\big\{ \{O_1,\dots,O_j\}\big\}$. 

\begin{figure}[t]\center
	\includegraphics[width=.2\textwidth]{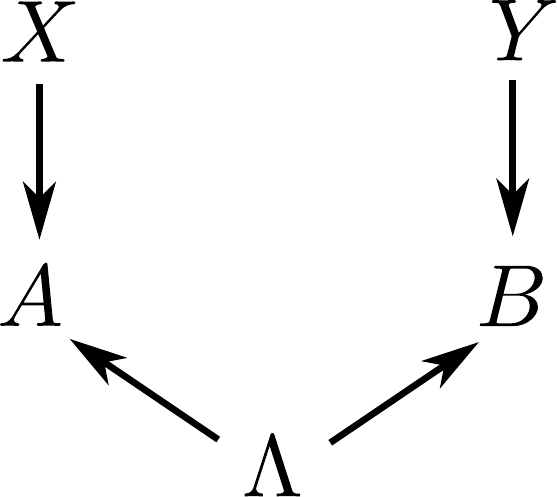}
	\caption{DAG showing the causal structure of a local hidden variable model for the Bell scenario. \label{fig:bell}}
\end{figure}

As an illustration, consider the paradigmatic causal Bayesian network for a local hidden variable model satisfying Bell's assumption of local causality~\cite{Bell1964,Wood2015}. 
The relevant DAG, shown on Fig.~\ref{fig:bell}, has five variables, four of which are observable while the hidden variable $\Lambda$ is not: in the context of Bell's Theorem the ``hidden variables'' indeed refer to the latent factors introduced above. 
This DAG represents the physical scenario where two distant observers receive physical systems produced by a common source (the hidden variable $\Lambda$) and make different measurements (choices of which are labelled by $X$ and $Y$), obtaining measurement outcomes (represented by the variables $A$ and $B$). 
That is, the probability structure is $\mathcal{S}=\{S\}$ with $S=\{X,Y,A,B,\Lambda\}$, and the marginal scenario is $\mathcal{M}=\big\{ \{X,Y,A,B\}\big\}$.
Some of the conditional independences implied by this DAG are given by $P(xy\lambda)=P(x)P(y)P(\lambda)$ (the measurement independence assumption), $P(a\vert xyb\lambda)=P(a\vert x\lambda)$ and $P(b\vert xya\lambda)=P(b\vert y\lambda)$ (the locality assumption) that in turn imply (after eliminating the hidden variable $\Lambda$) Bell inequalities for the observed variables~\cite{Bell1964,Wood2015}.
These constraints also imply the no-signaling constraints $P(a\vert xy)=P(a\vert x)$ and $P(b\vert xy)=P(b\vert y)$.

This example shows that, in general, DAGs with latent variables imply CIs both on the level of observable and unobservable variables. 
The CIs involving latent variables are not directly testable but imply further constraints (Bell inequalities, in the example above) that can be tested to check whether the observable behavior is compatible with the proposed underlying DAG.

If, instead of characterizing the allowed probability distributions, we consider the entropic description of the Bell scenario, i.e., the Shannon cone together with the linear constraints arising from the DAG's CIs, then after eliminating the latent variable $\Lambda$ one obtains no further constraints other than the elemental inequalities (which are trivial since they are respected by all probability distributions) and the observable CIs implied by the DAG: $H(XY)=H(X)+H(Y)$, $H(A\vert XY)=H(A\vert X)$ and $H(B\vert XY)=H(B\vert Y)$~\cite{Weilenmann:2016aa}. 
The first CI relation represents the independence of the two measurement choices, while the two latter ones are no-signaling conditions.
Thus, for this particular causal Bayesian network, when the entropic approach is applied to the variables making up the DAG one does not obtain any nontrivial constraints (i.e., entropic Bell inequalities)~\cite{Weilenmann:2016aa}. 
However, there are many examples of Bayesian networks for which one does obtain such nontrivial constraints~\cite{Chaves2014,Chaves2014b,Henson2014}. 
In fact, as we will see in Sec.~\ref{sec:bipartite}, a slight modification of this method also leads to nontrivial constraints on causal correlations.

\subsubsection{The entropic characterization of counterfactuals}
\label{sec:counterfactuals}

While the DAG method fails to provide nontrivial constraints for the Bell scenario (a result that can be extended to a larger class of ``line-like'' Bayesian networks~\cite{Weilenmann:2016aa}), it has been known for some time that entropic Bell inequalities can be derived using different methods~\cite{Braunstein1988}.
Interestingly, these inequalities can even be turned into necessary and sufficient conditions for a given probability distribution to satisfy Bell's local causality assumption~\cite{Chaves:2013aa}.

The method that allows such inequalities to be derived is motivated by the realization that the entropic approach can be applied to any marginal scenario for a relevant set of random variables~\cite{Fritz:2013aa}, and not only those arising from causal Bayesian networks. 
In particular, when we are interested in constraints on conditional distributions of the form $P(ab|xy)$, where we have distinct sets of input and output variables, we may consider the output variables conditioned on certain relevant input variables (e.g.\ $A_{xy}$ and $B_{xy}$, where the notation $A_{xy}$ denotes the random variable $A|(X=x,Y=y)$).\footnote{We focus here on the bipartite case for concreteness, but the method readily generalizes to multipartite scenarios.}
The choice of relevant input variables to condition on, as well as the appropriate probability structure, will depend on the physical situation being considered.
In general, a global probability distribution may not exist on such ``counterfactual'' variables even if one does exist on the unconditioned variables.

Let us illustrate how this method may be applied by considering again its application to the Bell scenario.
Instead of considering all the input and output variables as in the DAG approach (e.g.\ $X,Y,A,B$), one can consider copies of the output variables conditioned on the corresponding party's input, i.e., $A_{x},B_{y}$, where $A_{x}$ denotes the random variable $A{|(X=x)}$. 
Indeed, due to the no-signaling constraints, the output variables can only depend on the corresponding local input.
Furthermore, from Fine's Theorem~\cite{Fine1982} we know that Bell's local causality assumption is equivalent to the existence of a well defined (although empirically inaccessible) joint probability distribution $P(a_1,\dots, a_{\vert \cal X \vert},b_1,\dots, b_{\vert \cal Y \vert})$ (where ${\cal X} = \{1,\ldots,|{\cal X}|\}$ and ${\cal Y} = \{1,\ldots,|{\cal Y}|\}$ denote the alphabets of Alice and Bob's inputs) on these variables\footnote{In particular, by invoking Fine's Theorem we do not need to explicitly include the hidden variable $\Lambda$ in this method, contrary to the DAG method outlined previously.} that marginalizes to the observable one given by $P(ab \vert xy)=P(a_x,b_y)$. 
Hence, the appropriate probability structure for local correlations in the Bell scenario is $\mathcal{S}=\{S \}$ with $S=\{A_1\dots,A_{|\mathcal{X}|},B_1,\dots,B_{|\mathcal{Y}|}\}$, and we consider the Shannon cone $\Gamma_{S}=\Gamma^{\mathcal{S}}$ that contains all $2^{|\mathcal{X}|+|\mathcal{Y}|}$-dimensional entropy vectors $\bm{h}=\big(H(\emptyset), H(A_1),\dots, H(B_1),\dots, H(A_1\dots A_{\vert \cal X \vert}B_1\dots B_{\vert \cal Y \vert}) \big)$.
The marginal scenario in this case is simply $\mathcal{M}=\big\{ \{A_x,B_y\} \big\}_{x,y}$ and local correlations are then characterized by the cone $\Pi_\mathcal{M}\left(\Gamma_S\right)$. 

In contrast to the characterization based directly on the DAG variables, this approach leads to nontrivial entropic inequalities (i.e., not obtainable from the elemental inequalities in Eqs.~\eqref{shannonineqs_basic}) in the Bell scenario. 
For example, for two measurement settings per party, which we label in this case $x,y=0,1$, one obtains the Braunstein-Caves inequality~\cite{Braunstein1988} together with its symmetries obtained by relabeling the inputs, namely,
\begin{align}\label{eq:entropicBellIneq}
	 I(A_0:B_0)&+I(A_0:B_1) +I(A_1:B_0) \notag\\ 
	& -I(A_1:B_1) -H(A_0)-H(B_0) \leq 0 ,
\end{align}
where $I(A_x:B_y)\coloneqq H(A_x)+H(B_y)-H(A_xB_y)$ is the mutual information between the variables $A_x$ and $B_y$. 
This inequality can be understood as the entropic counterpart of the paradigmatic CHSH inequality~\cite{Clauser1969}. 

Although the choice of probability structure above corresponds, via Fine's theorem, to the assumption of a local hidden variable theory, one can also consider other possibilities.
For instance, taking $\mathcal{S}=\mathcal{M}$ amounts to assuming a nonsignaling theory~\cite{Popescu1994}. 
In this case, the entropy cone is characterized only by the Shannon inequalities and one can obtain a characterization of the extremal rays of the cone, corresponding to the entropic analogue of Popescu-Rohrlich (PR) boxes~\cite{Chaves:2016aa}. 

In general (i.e., beyond the simplest Bell scenario), both methods based on the variables in a causal Bayesian network and on counterfactual variables can lead to nontrivial constraints~\cite{Fritz2012,Chaves2014,Chaves2014b,Henson2014,Steudel2015,Chaves:2016aa}.
To conclude this section, let us nonetheless highlight an important difference between the two methods: while the former is valid for arbitrary input alphabets, the latter fixes the number of inputs to which the inequalities apply.

\section{Bipartite entropic causal inequalities}
\label{sec:bipartite}

With the entropic approach to characterizing sets of correlations outlined, we can now proceed to apply this approach to causal correlations, so as to derive \emph{entropic causal inequalities}.
We consider in this section the bipartite case. We first show how the method based on causal Bayesian networks can be adapted to characterize causal correlations, before considering also the method based on counterfactual variables.

\subsection{Characterization based on causal Bayesian networks}
\label{sec:causal_str_bi}

\subsubsection{Conditional DAGs for bipartite causal correlations}

The ability to apply the entropic approach to DAGs, as outlined in Sec.~\ref{sec:causal_str}, is a powerful tool for characterizing the correlations obtainable within arbitrary causal networks.
However, the notion of causal correlations defined in Eq.~\eqref{eq:causalCorreltn} is somewhat more general and cannot be directly expressed within the framework of causal Bayesian networks.
In order to see why this is the case, let us first note that the random variables of interest are $X,Y,A,B$, representing the inputs $X,Y$ and outputs $A,B$ for Alice and Bob.
Note that since we consider signaling scenarios here, unlike in the Bell scenario, we do not need to include any latent variable $\Lambda$ in our description to account for shared randomness, since this can be established via local randomness and communication.

\begin{figure}[t]\center
\includegraphics[width=.25\textwidth]{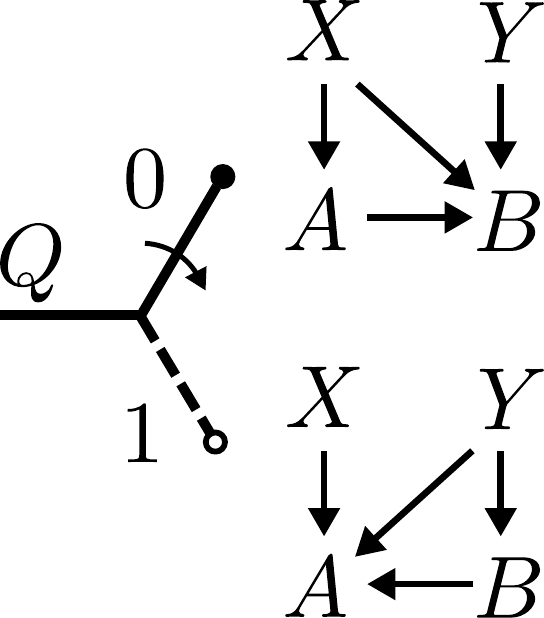}
\caption{DAGs for bipartite causal correlations. The latent ``switch'' variable $Q$ determines which DAG, corresponding to the fixed causal order $\mathrm{A}\prec\mathrm{B}$ (for $Q=0$, top) or $\mathrm{B}\prec\mathrm{A}$ (for $Q=1$, bottom), is ``activated''.\label{fig:dag_bi}}
\end{figure}

If Alice and Bob share a correlation compatible with a fixed causal order (i.e.\ either $\mathrm{A}\prec \mathrm{B}$ or $\mathrm{B}\prec \mathrm{A}$), then the functional dependences between $X,Y,A,B$ can indeed be expressed as a DAG -- specifically, the two DAGs containing these variables in Fig.~\ref{fig:dag_bi}.
However, a causal correlation may in general not be compatible with any fixed causal order, but may require a mixture thereof. 
This has some similarities with the situation in the Svetlichny definition of genuine multipartite nonlocality~\cite{Svetlichny1987,Chaves2016causal} where a convex mixture of different DAGs has to be considered.

To tackle this problem it is necessary to find a way to take into account the constraints arising separately from each of the two fixed causal orders, and then to combine them  to obtain those satisfied by causal correlations.
In order to do this, we exploit the fact that any mixture of fixed-order causal correlations can be seen as arising from a latent variable that determines the causal order for each individual experiment~\cite{oreshkov15}. We thus introduce a new random variable $Q$ which we call a ``switch'', and which determines univocally the appropriate causal Bayesian network for each trial.
The resulting causal model is shown in Fig.~\ref{fig:dag_bi}, where the DAG with $\mathrm{A}\prec\mathrm{B}$ is used for $Q=0$, and the one with $\mathrm{B}\prec\mathrm{A}$ for $Q=1$. By identifying $q_0,q_1$ in Eq.~\eqref{eq:causalCorreltn} as $q_0=P(Q=0)$, and $q_1=P(Q=1)$, one can readily see that this description is equivalent to the definition of causal correlations in Eq.~\eqref{eq:causalCorreltn}.

Both DAGs imply the independence of the inputs, $X\CI Y$. The DAG for $Q=0$ (i.e., for $\mathrm{A}\prec\mathrm{B}$) also implies the CI $A\CI Y\mid X$ (i.e.\ that there is no signaling from $\mathrm{B}$ to $\mathrm{A}$), while the DAG for $Q=1$ implies $B\CI X\mid Y$ instead.
In addition, the switch variable $Q$ should be independent of Alice and Bob's inputs $X$ and $Y$, so that we have $XY\CI Q$, which, together with $X\CI Y$, implies that $X\CI Y\CI Q$.

\subsubsection{Shannon polyhedra of causal correlations}\label{sec:Sh_pol_dag}

In order to use the ``conditional'' causal Bayesian network in Fig.~\ref{fig:dag_bi} to characterize the set of entropy vectors obtainable from causal correlations, we first note that we can directly use the techniques of Sec.~\ref{sec:causal_str} to construct the Shannon cones for each of the two DAGs appearing in the figure conditioned on $Q$ (i.e., for fixed-order correlations with $\mathrm{A}\prec\mathrm{B}$ or $\mathrm{B}\prec\mathrm{A}$).
Denoting these cones $\Gamma^{\mathrm{A}\prec\mathrm{B}}$ and $\Gamma^{\mathrm{B}\prec\mathrm{A}}$, we have
\begin{equation}\label{eq:coneAB}
	\Gamma^{\mathrm{A}\prec\mathrm{B}}=\Gamma_{S}\cap\textrm{L}_{\mathcal{C}}^{\mathrm{A}\prec\mathrm{B}}
\end{equation}
and 
\begin{equation}\label{eq:coneBA}
	\Gamma^{\mathrm{B}\prec\mathrm{A}}=\Gamma_{S}\cap\textrm{L}_{\mathcal{C}}^{\mathrm{B}\prec\mathrm{A}},
\end{equation}
where $\Gamma_{S}$ is the Shannon cone for the four variables in $S=\{X,Y,A,B\}$, the probability structure is simply $\mathcal{S}=\{S\}$, and $\textrm{L}_{\mathcal{C}}^{\mathrm{A}\prec\mathrm{B}}$ denotes the linear subspace defined by the CI constraints for the case ${\mathrm{A}\prec\mathrm{B}}$, namely, the equations $H(XY)=H(X)+H(Y)$ and $H(YA| X)= H(Y| X) + H(A|X)$, and similarly for $\textrm{L}_{\mathcal{C}}^{\mathrm{B}\prec\mathrm{A}}$.
These cones are characterized by the systems of inequalities $I_0\bm{h}\le \bm{0}$ and $I_1\bm{h}\le \bm{0}$, where $\bm{h}=(H(T))_{T\subset S}$. 

Recall that in the probabilistic case the polytope of causal correlations is simply the convex hull of the polytopes of correlations for $\mathrm{A}\prec\mathrm{B}$ and $\mathrm{B}\prec\mathrm{A}$~\cite{Branciard:2016aa}, and with the new variable $Q$ the definition in Eq.~\eqref{eq:causalCorreltn} can be rewritten as
\begin{align}\label{eq:causalCorreltn_Q}
	P(ab|xy)=& \ P(Q=0)P^{\mathrm{A}\prec\mathrm{B}}(ab|xy,Q=0) \notag \\
&+P(Q=1)P^{\mathrm{B}\prec\mathrm{A}}(ab|xy,Q=1).
\end{align}
In contrast, the convex hull of the cones $\Gamma^{\mathrm{A}\prec\mathrm{B}}$ and $\Gamma^{\mathrm{B}\prec\mathrm{A}}$ does not contain all entropy vectors of causal correlations due to the concavity of the Shannon entropy.
Indeed, in Appendix~\ref{apndx:counterex} we provide an explicit example of a causal correlation whose entropy vector is not contained in the convex hull $\conv(\Gamma^{\mathrm{A}\prec\mathrm{B}},\Gamma^{\mathrm{B}\prec\mathrm{A}})$.

To see more precisely why this is the case, and how to give a correct entropic characterization of causal correlations, observe that, when taking a convex mixture of two causal correlations with different causal orders, the ``conditional entropy vectors'' $\bm{h}_0=(H(T|Q=0))_{T\subset S}$ and $\bm{h}_1=(H(T|Q=1))_{T\subset S}$ must be contained in $\Gamma^{\mathrm{A}\prec\mathrm{B}}$ and $\Gamma^{\mathrm{B}\prec\mathrm{A}}$, respectively, and thus satisfy $I_0\bm{h}_0\le\bm{0}$ and $I_1\bm{h}_1\le\bm{0}$.
For any causal correlation, the convex mixture
\begin{equation}\label{eq:convHvec}
	\bm{h}_{\textrm{conv}} = P(Q=0)\bm{h}_0 + P(Q=1)\bm{h}_1
\end{equation}
is thus contained in $\conv(\Gamma^{\mathrm{A}\prec\mathrm{B}},\Gamma^{\mathrm{B}\prec\mathrm{A}})$.
Observe now that, in contrast to the convex sum Eq.~\eqref{eq:causalCorreltn_Q} defining causal correlations, $\bm{h}_{\textrm{conv}}$ thus defined is equal to $(H(T|Q))_{T\subset S}$, rather than just $(H(T))_{T\subset S}$, and hence the convex hull of the fixed-order cones characterizes the conditional entropies (conditioned on the switch variable $Q$) obtainable with causal correlations, rather than the entropy vectors of causal correlations directly.

With the appropriate transformation, the inequalities $I\bm{h}\le\bm{0}$ characterizing\footnote{In practice these can be obtained by taking the union of the extremal rays of the two cones $\Gamma^{\mathrm{A}\prec\mathrm{B}}$ and $\Gamma^{\mathrm{B}\prec\mathrm{A}}$ and solving the facet enumeration problem to obtain the inequality representation of $\conv(\Gamma^{\mathrm{A}\prec\mathrm{B}},\Gamma^{\mathrm{B}\prec\mathrm{A}})$ using standard software for convex polyhedra such as PANDA~\cite{PANDA}.} 
$\conv(\Gamma^{\mathrm{A}\prec\mathrm{B}},\Gamma^{\mathrm{B}\prec\mathrm{A}})$ can be transformed into inequalities satisfied by the standard (i.e., non-conditional) entropy vector $\tilde{\bm{h}}=(H(T))_{T\subset \tilde{S}}$ for the variables now in $\tilde{S}=S\cup\{Q\}$ (and the probability structure is consequently extended to $\widetilde{\mathcal{S}}=\{\tilde{S}\}$).
Specifically, each row $\bm{I}$ of the matrix $I$ (defining each individual inequality $\bm{I}\cdot\bm{h}\le 0$) must undergo the linear transformation $\mathcal{T}_{Q}: \mathds{R}^{2^{|S|}}\rightarrow \mathds{R}^{2^{|S|+1}}$ mapping $\bm{I}\mapsto\tilde{\bm{I}}\coloneqq \mathcal{T}_{Q}(\bm{I})$ with the components of $\tilde{\bm{I}}$ given by\footnote{Note that $(\tilde{\bm{I}})_{\emptyset}$ multiplies $H(\emptyset) = 0$ in the scalar product $\tilde{\bm{I}}\tilde{\bm{h}}$, so its value is irrelevant.}
\begin{equation}\label{eq:lin_tran}
	(\tilde{\bm{I}})_{T \cup \{Q\}} = (\bm{I})_T, \ (\tilde{\bm{I}})_{\{Q\}} = -\sum_{T \neq \emptyset} (\bm{I})_T,  \text{ and } (\tilde{\bm{I}})_T=0
\end{equation}
for all nonempty subsets  $T\subset S$.
We will denote by $\conv_Q(\Gamma^{\mathrm{A}\prec\mathrm{B}},\Gamma^{\mathrm{B}\prec\mathrm{A}})$ the cone of vectors $\tilde{\bm{h}}$ satisfying the resulting inequalities $\tilde{I}\tilde{\bm{h}}\le\bm{0}$.

To complete the characterization of entropy vectors for causal correlations, we recall that, in addition to the fact that any distribution on $\tilde{S}$ must give an entropy vector in the Shannon cone $\Gamma_{\tilde{S}}$, the conditional DAG in Fig.~\ref{fig:dag_bi} gives us the CI constraints $X\CI Y\CI Q$. Moreover, since  $Q$ is a binary variable (as there are only two orders to switch between) we have $H(Q)\le 1$. 
A consequence of this final inequality constraint is that the set of entropy vectors under consideration will be characterized by an inhomogeneous system of inequalities of the form $\tilde{I}\tilde{\bm{h}}\le \tilde{\bm{\beta}}$ for some $\tilde{\bm{\beta}}\in\mathbb{R}^{2^{|S|+1}}$ and is thus no longer a cone but a polyhedron.
The polyhedron characterizing entropy vectors associated with the conditional DAG (when still including $Q$) is thus given by
\begin{align}\label{eq:bip_cau_coneQ}
	 \widetilde\Gamma^{\mathrm{causal}}_{\rm AB} =& \Gamma_{\tilde{S}} \cap \conv_Q(\Gamma^{\mathrm{A}\prec\mathrm{B}},\Gamma^{\mathrm{B}\prec\mathrm{A}}) \notag\\
	& \cap \textrm{L}_{\mathcal{C}}\big(\{(X\CI Y\CI Q), H(Q)\leq 1\}\big),
\end{align}
where the notation $\textrm{L}_{\mathcal{C}}(\cdot)$ denotes the subset (here, a polyhedron) in the entropy vector space defined by the corresponding linear constraints.

Finally, following the general approach presented in Sec.~\ref{sec:entropymarginal}, it remains just to eliminate the terms containing the (unobservable) switch variable $Q$ in order to obtain the inequalities characterizing bipartite causal correlations.
This is done by projecting $\widetilde\Gamma^{\mathrm{causal}}_{\rm AB}$ onto the marginal scenario $\mathcal{M}= \{S\} = \big\{ \{X,Y,A,B\}\big\}$.
We thus finally obtain the polyhedron
\begin{align}\label{eq:bip_cau_cone}
	&\Gamma^{\mathrm{causal}}_{\rm AB} = \Pi_{\mathcal{M}} \big( \widetilde\Gamma^{\mathrm{causal}}_{\rm AB} \big),
\end{align}
which we shall refer to as the \emph{causal Shannon polyhedron} or simply the \emph{causal polyhedron} and is again characterized by an inhomogeneous system of inequalities $I'\bm{h}\le \bm{\beta}$ for some $\bm{\beta}\in\mathbb{R}^{2^{|S|}}$.

We emphasize that the construction given above is in fact not at all restricted to the description of causal correlations, and can be used to characterize arbitrary convex mixtures of different Bayesian networks. Furthermore, as we will see in Sec.~\ref{sec:multipartite}, this method can be generalized to convex combinations of more distributions, in our case corresponding to more than two causal orders in multipartite scenarios (and even correlations with ``dynamical causal order''~\cite{hardy2005probability,oreshkov15,Abbott:2016aa}).

\subsubsection{Entropic causal inequalities and their violation}\label{sec:DAGviolations}

The constructive description of the causal polyhedron $\Gamma^{\textrm{causal}}_{\rm AB}$ from Eqs.~\eqref{eq:bip_cau_coneQ} and~\eqref{eq:bip_cau_cone} also makes it clear how we can characterize it, in practice, as a system of linear inequalities.
A description of $\widetilde\Gamma^{\mathrm{causal}}_{\rm AB}$ in terms of its facets is straightforwardly obtained by taking the union of the inequalities describing the individual cones and polyhedron appearing in Eq.~\eqref{eq:bip_cau_coneQ} and eliminating redundant ones.
The inequalities characterizing $\Gamma^{\textrm{causal}}_{\rm AB}$ can then be found by eliminating the terms not contained in the marginal scenario $\mathcal{M}= \{S\}$, either by Fourier-Motzkin elimination~\cite{Williams1986} or by finding its extremal rays and projecting out the unwanted coordinates.

The resulting system of inequalities is thus satisfied by any bipartite causal correlation.
However, many of these inequalities are either elemental inequalities (as in Eq.~\eqref{shannonineqs_basic}) or can be obtained from these by using the independence constraint $X\CI Y$, and thus represent trivial constraints.
After characterizing the polyhedron in Eq.~\eqref{eq:bip_cau_cone} and eliminating all trivial inequalities, i.e., those satisfied by any distribution $P(xyab)$ with $X\CI Y$, we find 35 novel entropic causal inequalities.
Several of these inequalities are equivalent under the exchange of parties (i.e., exchanging $(X,A)\leftrightarrow(Y,B)$), and under this symmetry there are in fact 20 equivalence classes of entropic causal inequalities, the full list of which is given in Appendix~\ref{apndx:DAGineqs}.
Of these, 10 have bounds of $0$ (i.e., are of the form $\bm{I}\cdot\bm{h}\le 0$), while the remaining 10 have nonzero bounds (resulting from a nontrivial dependence on $H(Q)$ before this variable was eliminated; see Appendix~\ref{apndx:DAGineqs}).
Simple interpretations of the entropic causal inequalities seem to be less forthcoming than for the bipartite causal inequalities in terms of probabilities~\cite{Branciard:2016aa} (for binary inputs and outputs  -- recall that the entropic inequalities given here are, in contrast, valid for any number of possible inputs and outputs).
One of the simpler examples, which is symmetric under the exchange of parties, is
\begin{equation}\label{eq:dagineq1}
	I(X:YA) + I(Y:XB) - H(AB) \leq 0.
\end{equation}
Note that the fact that we find nontrivial inequalities is in stark contrast to the situation for Bell-type inequalities (and line-like causal Bayesian networks), where the DAG-based entropic method only leads to trivial inequalities obtainable from the elemental inequalities and no-signaling conditions~\cite{Weilenmann:2016aa}.

While these entropic inequalities are obeyed by any bipartite causal correlation, we note that \emph{a priori} they need not be tight.
Indeed, recall that the Shannon cone is only an outer approximation to the true entropy cone, so it is thus interesting to study the tightness and violation of these inequalities more carefully.

Although one generally would not expect every point on the boundary of $\Gamma^{\mathrm{causal}}_{\rm AB}$ to be obtainable by a causal correlation, it is nonetheless desirable to be able to saturate each inequality by some causal probability distribution for appropriate distributions for $X$ and $Y$.
By looking at deterministic causal distributions with binary inputs and outputs, which can easily be enumerated, we readily verified that all 10 families of inequalities that are bounded by $0$ (given in Eq.~\eqref{eq:dag_v1}) can indeed be saturated when taking uniformly distributed inputs.
However, we were unable to find causal distributions, either by mixing binary ones or by considering more outputs, that saturate the remaining inequalities, and their tightness remains an open question.

To understand now the violation by noncausal distributions of the entropic inequalities, we consider the extremal rays of the constrained Shannon cone
\begin{equation}\label{eq:constrainedShannonCone}
	\Gamma_S\cap\textrm{L}_{\mathcal{C}}\big( \{X\CI Y\} \big)
\end{equation}
which violate the inequalities.\footnote{Note that the nontriviality of the inequalities implies that such extremal rays indeed exist.}
A crucial question is whether or not these extremal rays actually correspond to valid probability distributions (i.e., whether they support entropy vectors), and if not, whether the inequalities can nonetheless be violated.

In order to look at this, it is instructive to first restrict our attention to distributions satisfying $H(X)\le 1$, $H(Y)\le 1$, $H(A)\le 1$ and $H(B)\le 1$.
These constraints are satisfied by all distributions with binary inputs and outputs, and this therefore also allows us to compare the violation of the entropic causal inequalities to the violation of standard causal inequalities that are understood well in this scenario~\cite{Branciard:2016aa}.
Imposing these constraints on the cone in Eq.~\eqref{eq:constrainedShannonCone}, one obtains a polytope with extremal points corresponding to the extremal rays of the cone scaled to satisfy these constraints (together with the null vertex $\bm{0}$).
Under these we found that the 10 inequalities in Eq.~\eqref{eq:dag_v1} and the two inequalities in Eq.~\eqref{eq:dag_v2} could be violated, although the latter are (in that case) weaker than, and implied by, the former and are thus redundant.
The remaining 8 inequalities in Eqs.~\eqref{eq:dag_v3} and~\eqref{eq:dag_v4} cannot be violated. 
All in all,  the set of binary causal correlations is entropically characterized by the 10 inequalities in Eq.~\eqref{eq:dag_v1} that are bounded by $0$.

Amongst the extremal points violating each of these inequalities, those that give the maximal violation all satisfy $H(X)=H(Y)=1$ and  $H(XY)=H(XYAB)$ and thus, if realizable, correspond to deterministic conditional distributions taken with uniformly distributed inputs $X$ and $Y$.%
\footnote{Note that there are nonetheless extremal points that are only realizable by distributions with non-uniformly distributed inputs $X$ and $Y$, and which violate some of the inequalities. However, such distributions never yield the maximal violation obtainable.}
In fact, all but one of these 10 inequalities are maximally violated (by which we henceforth mean with respect to the Shannon cone augmented with the independence constraint $X \CI Y$) by one of the three following deterministic distributions taken with uniform inputs:
\begin{align}\label{eq:GYNIdists}
	P(ab|xy) &= \delta_{a,y}\,\delta_{b\oplus y,x} \notag\\
	P(ab|xy) &= \delta_{a\oplus x, y}\,\delta_{b,x}\\
	P(ab|xy) &= \delta_{a\oplus x, y}\,\delta_{b\oplus y,x}, \notag
\end{align}
where $x,y,a,b$ take the binary values $0,1$, and $\oplus$ denotes addition modulo $2$.
For example, Eq.~\eqref{eq:dagineq1} is violated by the third distribution above with a value for the left-hand side of $1$.
The one exception not violated by the distributions in Eq.~\eqref{eq:GYNIdists} is the second inequality in~\eqref{eq:dag_v1},
\begin{align}
	 I(A:B) &- I(A:B|X) - I(A:B|Y) - 2H(AB|XY) \leq 0,
	 \label{eq:dagineq2}
\end{align}
which, in turn, is violated by the deterministic distribution (again taken with uniform inputs)
\begin{equation}\label{eq:otherdist}
	P(ab|xy) = \delta_{a\oplus x,xy}\,\delta_{b\oplus y,xy}.
\end{equation}
However, unlike for the other inequalities, this distribution does not give the maximal possible violation of inequality~\eqref{eq:dagineq2} (which is $1/2$), 
as the corresponding extremal point $\bm{h}_{\mathrm{ext}}$ that does maximally violate it is not reachable by a valid probability distribution with binary inputs and outputs. 
This is easily verified by making use of the previous observation that this extremal point must correspond to a deterministic distribution taken with uniform inputs, the set of which can easily be enumerated for binary inputs and outputs.
Amongst such distributions, the one in Eq.~\eqref{eq:otherdist} gives the best violation of $1-\frac{3}{2}\log_2\frac{3}{2}\approx 0.123 > 0$. 

The distributions in Eq.~\eqref{eq:GYNIdists} are particularly interesting, as they all violate maximally some symmetries of the GYNI inequality~\eqref{eq:GYNI} (under relabeling of the inputs and outputs), but not Eq.~\eqref{eq:GYNI} itself.
Interestingly, it turns out that \emph{all} binary deterministic noncausal distributions, when taken with uniform inputs, violate at least one of our entropic inequalities \emph{except} the distribution $P^{\mathrm{GYNI}}(ab|xy) = \delta_{a,y}\delta_{b,x}$ (which violates maximally Eq.~\eqref{eq:GYNI}) and its four symmetries under input-independent relabeling of outputs only.
Note, however, that if Alice and Bob have a noncausal resource producing the distribution $P^{\mathrm{GYNI}}$, they can produce any of the distributions in Eq.~\eqref{eq:GYNIdists} by appropriately XORing their input with their output, and thus still obtain an operational violation of an entropic causal inequality.\footnote{This illustrates an important difference between the probabilistic and entropic frameworks: while all symmetries of a correlation obtained by flipping inputs and outputs (possibly conditioned on the local inputs for the latter) are equivalent in the probabilistic case (in the sense that if one violates a causal inequality, then all other ones violate a symmetry of that inequality) this is not the case in the entropic approach. The entropy vectors of two different symmetries of a correlation may be inequivalent, with one violating an entropic causal inequality while the other does not.}
It is interesting to observe that distributions maximally violating GYNI-type inequalities have such a crucial role in violating the entropic causal inequalities given that the entropic inequalities superficially bear little resemblance to these, and are valid for arbitrary numbers of inputs and outputs.

Returning to the more general situation with no upper bound imposed on $H(X), H(Y), H(A)$ and $H(B)$, we see that all the remaining entropic causal inequalities can be violated by entropy vectors that are parallel to the realizable entropy vectors giving violations in the restricted scenario -- more precisely, those obtained from the distributions Eq.~\eqref{eq:GYNIdists} (for all but one of the remaining inequalities) and Eq.~\eqref{eq:otherdist} (for the remaining one). 
This shows that, given large enough alphabets for the input and output variables, all the entropic causal inequalities we obtained can indeed be violated by noncausal probability distributions, since if the distribution $P(xyab)$ has entropy vector $\bm{h}$ then the distribution 
\begin{equation}\label{eq:mult_cp_nc}
	P(\bm{x}\bm{y}\bm{a}\bm{b})=P(x_1y_1a_1b_1)\times \cdots\times P(x_ny_na_nb_n),
\end{equation}
where $\bm{x}=(x_1,\dots,x_n)$ and similarly for $\bm{y}$, $\bm{a}$ and $\bm{b}$, has entropy vector $n\cdot\bm{h}$.%
\footnote{Using this approach (and the distributions in Eqs.~\eqref{eq:GYNIdists} and~\eqref{eq:otherdist}), one must take $n=17$ (and thus $2^{17}$ inputs) in order to violate all the remaining inequalities (although $n=2$ is sufficient for all but one of these inequalities). However, we expect that more intelligent approaches may allow them to be violated using less inputs.}
One should be careful, however, to note that the operation of sharing multiple independent correlations among the same parties is not a free operation either in the framework of causal correlations (since, for example, two independent copies of a causal distribution may give rise to a noncausal one), or in the process matrix framework (where two independent copies of a process matrix does not, in general, produce a valid process matrix). 
Nevertheless, $P(\bm{a}\bm{b}|\bm{x}\bm{y}) = P(\bm{x}\bm{y}\bm{a}\bm{b})/P(\bm{x}\bm{y})$ obtained from Eq.~\eqref{eq:mult_cp_nc} still represents a valid (possibly noncausal) distribution.

It is interesting also to ask how sensitive the entropic causal inequalities are for detecting noncausality.
Since it does not appear possible to saturate the inequalities~\eqref{eq:dag_v2}--\eqref{eq:dag_v4} with non-zero bounds using causal distributions, these inequalities are not tight and, consequentially, unable to detect noncausal correlations that are very close to being causal.
For the other inequalities in Eq.~\eqref{eq:dag_v1} this is nonetheless a pertinent question.
More precisely, one may ask whether there exists a distribution $P^\varepsilon$ of the form
\begin{equation}
\label{eq:mix_distr}
	P^\varepsilon(ab|xy)= \varepsilon P^{\textrm{NC}}(ab|xy) + (1-\varepsilon)P^{\textrm{C}}(ab|xy),
\end{equation}
where $P^{\textrm{NC}}$ is a noncausal distribution and $P^{\textrm{C}}$ is causal, that violates any of these entropic inequalities for arbitrarily small $\varepsilon>0$.

We looked in detail at this question for the case of binary inputs and outputs, where the inequalities in Eq.~\eqref{eq:dag_v1} can all both be saturated by causal distributions, and violated by noncausal ones.
By trying exhaustively all deterministic distributions $P^{\textrm{NC}}$ and $P^{\textrm{C}}$, we found that such behaviour was exhibited (for such distributions) only by the two inequalities
\begin{equation}\label{eq:goodmixingineq1}
	I(A:B|X) - I(Y:B) - 2H(B|XY) \leq 0
\end{equation}
and 
\begin{equation}\label{eq:goodmixingineq2}
	I(XA:Y) + I(YB:X) - H(X|YA) - H(A) \leq 0.
\end{equation}
Equation~\eqref{eq:goodmixingineq1}, for example, is violated by $P^\varepsilon$ for all $\varepsilon>0$ when taking $P^{\textrm{NC}}(ab|xy)=\delta_{a\oplus x,y}\,\delta_{b\oplus y,x}$ and $P^{\textrm{C}}(ab|xy)=\delta_{a,0}\,\delta_{b\oplus y,x}$ along with uniformly distributed inputs $X$ and $Y$, which also gives a violation of the GYNI-type causal inequality 
\begin{equation}
	\frac{1}{4}\sum_{x,y,a,b} \delta_{a\oplus x,y}\, \delta_{b\oplus y,x}\, P(ab|xy)\le\frac{1}{2}
\end{equation}
with a left-hand side value of $\frac{1+\varepsilon}{2} > \frac{1}{2}$.

For the remaining inequalities, such mixtures that violate a standard causal inequality for arbitrarily small $\varepsilon$ only violate an entropic causal inequality when $\varepsilon>\varepsilon_0$ for some $\varepsilon_0$ bounded away from $0$. 
We observed identical behavior when we extended our consideration also to various non-deterministic distributions $P^{\textrm{NC}}$ and $P^{\textrm{C}}$, and it thus seems that only Eqs.~\eqref{eq:goodmixingineq1} and~\eqref{eq:goodmixingineq2} exhibit this ability to detect the noncausality of distributions that are arbitrarily close to being causal.

A final point worth discussing relates to the physical interpretation of the distributions violating entropic causal inequalities.
One of the motivations in introducing the notion of causal correlations was whether nature permits more general causal structures that might allow such correlations to be realized, for example in quantum gravity.
In particular, the authors of Ref.~\cite{oreshkov12} introduced the so-called process matrix formalism, in which quantum mechanics is assumed to hold locally for each party, while no global order is assumed between the parties.
They showed that causal inequalities can be violated within this framework, and this helped motivate further studies of causal and noncausal correlations, where it has been shown that the violation of causal inequalities is ubiquitous within this framework~\cite{Abbott:2016aa,baumeler13,baumeler14,Baumelerspace2016,Branciard:2016aa,feix16}.
It is thus interesting to see whether entropic causal inequalities share this property and can also be violated within the process matrix framework.

To look for such violations, we used the optimization techniques of Refs.~\cite{Branciard:2016aa,Abbott:2016aa} with qubit systems to try and optimize the violation of the GYNI-type inequalities that the distributions in Eq.~\eqref{eq:GYNIdists} violate maximally.
We also tried minimizing the distance to other deterministic noncausal correlations such as Eq.~\eqref{eq:otherdist}, as well as optimizing in random directions in probability space.
Unfortunately, we were unable to find any process matrices operating on qubits that violate our entropic causal inequalities with such techniques.
We additionally attempted to reproduce (as closely as possible) distributions of the form~\eqref{eq:mix_distr} for small $\varepsilon$ in order to violate inequalities~\eqref{eq:goodmixingineq1} and~\eqref{eq:goodmixingineq2}, but similarly found no violation.
Finally, we looked at correlations obtained by mixing noncausal correlations realizable by process matrices with causal correlations.
An analogous mixing procedure was shown to enable all nonlocal distributions to violate the entropic Bell inequalities described in Sec.~\ref{sec:counterfactuals}~\cite{Chaves:2013aa}, but we were unable to find violations of any entropic causal inequalities with this approach.

This lack of violation is perhaps unsurprising given the general lack of sensitivity of the entropic inequalities to nearly-causal distributions, and the fact that the best-known violations of causal inequalities for this scenario with process matrices are relatively small~\cite{Branciard:2016aa}. 
Nonetheless, it remains possible that violations can be found with higher-dimensional systems or more inputs and outputs; we leave this as an open question.

\subsection{Characterization based on counterfactual variables}
\label{sec:counterfactuals_causal}

In this section we will consider counterfactual variables as outlined in Sec.~\ref{sec:counterfactuals}.
Rather than considering the inputs as random variables $X$ and $Y$, we take copies of each output variable for all input combinations, i.e.~$A_{xy}$ and $B_{xy}$. 
In contrast to the method based on causal Bayesian networks, this method fixes the number of inputs that the inequalities apply to but may lead to novel constraints, as is the case in the Bell scenario.

\subsubsection{Counterfactual variables for bipartite causal correlations}

To keep the discussion simple, we will consider only the case of binary inputs, but the generalization to arbitrary inputs is straightforward.
We thus consider the variables in
\begin{equation}
	S = \{A_{00},A_{01},A_{10},A_{11},B_{00},B_{01},B_{10},B_{11}\}.
\end{equation}
Note that, in contrast to the example of Bell inequalities discussed in Sec.~\ref{sec:counterfactuals}, we need to consider copies of each variable for each input pair $(x,y)$.
This is a consequence of the fact that the correlations which we want to characterize may be signaling, e.g.,  for the causal order $\mathrm{A}\prec\mathrm{B}$, $B_{00}$ and $B_{10}$ will in general be different.

Since $A_{xy}$ and $B_{x'y'}$ are jointly observable only if $x=x'$ and $y=y'$, the marginal scenario in this case is
\begin{equation}\label{eq:counterfactualM}
	\mathcal{M}=\Big\{\{A_{00},B_{00}\},\{A_{01},B_{01}\},\{A_{10},B_{10}\},\{A_{11},B_{11}\}\Big\}.
\end{equation}
In contrast to the DAG-based method, several choices of probability structure $\mathcal{S}$ compatible with $\mathcal{M}$ are possible, and the particular choice must be motivated on the basis of physical assumptions.
One natural possibility would be to take $\mathcal{S}=\mathcal{M}$, as one may have no \emph{a priori} reason to think that the variables $A_{xy}$ and $A_{x'y'}$ have simultaneous physical meaning for $(x,y)\neq (x',y')$, and hence may not have a well-defined joint probability distribution.
On the other hand, in some cases one may imagine that such inputs correspond to the choice of measurements of some physical properties that are simultaneously well-defined, as in a classical theory; hence, one may alternatively take $\mathcal{S}= \{ \cup_{M_j\in \mathcal{M}} M_j\}=\{S\}$.
In the following, we will adopt the former approach and take $\mathcal{S}=\mathcal{M}$, since this constitutes the minimum assumptions compatible with the marginal scenario.
The Shannon cone for $\mathcal{S}$ is thus
\begin{equation}\label{eq:shannonCone_counterfact}
	\Gamma^{\mathcal{S}} = \Gamma_{\{A_{00},B_{00}\}}\cap\Gamma_{\{A_{01}B_{01}\}}\cap\Gamma_{\{A_{10}B_{10}\}}\cap\Gamma_{\{A_{11}B_{11}\}},
\end{equation}
as in Eq.~\eqref{eq:defconeS}.
We note however that this physically motivated choice for $\mathcal{S}$ implies, for this particular scenario, that a global probability distribution does in fact exist.\footnote{
This is the result of the more general fact that different choices of $\mathcal{S}$ may provide equivalent descriptions of marginal probabilities~\cite{Vorob1962} and entropies~\cite{BMC2016}.}
Taking $\mathcal{S}=\{S\}$ would thus provide an equivalent entropic characterization, and moreover, this equivalence also holds at the level of Shannon (rather than entropy) cones (see Appendix~\ref{sec:A_gp} for a more detailed discussion). 

We follow a method analogous to that used in Sec.~\ref{sec:causal_str_bi}.
First, we characterize the cones $\Gamma^{\mathrm{A}\prec\mathrm{B}}$ and $\Gamma^{\mathrm{B}\prec\mathrm{A}}$ of entropy vectors for fixed-order causal correlations, then, we characterize the convex mixtures of such correlations.

To do this, we note that the no-signaling conditions obeyed by fixed-order correlations (see Sec.~\ref{sec:causalcorr}) impose constraints on the counterfactual variables.
For example, correlations consistent with the order $\mathrm{A}\prec\mathrm{B}$ obey $P(a|xy)=P(a|xy')$ for all $x,y,y',a$, which implies $A_{xy}=A_{xy'}$ and thus $H(A_{xy})=H(A_{xy'})$ also.
Similarly, for $\mathrm{B}\prec\mathrm{A}$, we have $H(B_{xy})=H(B_{x'y})$ for all $x,x',y$.
The cones $\Gamma^{\mathrm{A}\prec\mathrm{B}}$ and $\Gamma^{\mathrm{B}\prec\mathrm{A}}$ are thus given by
\begin{equation}\label{eq:coneABcoutnerfact}
	\Gamma^{\mathrm{A}\prec\mathrm{B}} = \Gamma^\mathcal{S}\cap\textrm{L}_{\mathcal{C}}\big(\{A_{00}=A_{01},\; A_{10}=A_{11}\}\big)
\end{equation}
and
\begin{equation}\label{eq:coneBAcoutnerfact}
	\Gamma^{\mathrm{B}\prec\mathrm{A}} = \Gamma^\mathcal{S}\cap\textrm{L}_{\mathcal{C}}\big(\{B_{00}=B_{10},\; B_{01}=B_{11}\}\big),
\end{equation}
where $\textrm{L}_{\mathcal{C}}(\cdot)$ again denotes the linear subspace defined by the corresponding constraints.
 
As in Sec.~\ref{sec:causal_str_bi}, we introduce the latent switch variable $Q$, denote the augmented set of random variables $\tilde{S}=S\cup\{Q\}$, and extend the probability structure as
\begin{equation}\label{eq:tildeS}
	\widetilde{\mathcal{S}}=\Big\{ \{A_{xy},B_{xy},Q\} \mid x,y\in\{0,1\} \Big\}
\end{equation}
(in Appendix~\ref{sec:A_gp} we discuss further the implications of different choices of probability structures).
With this extra variable we note again that the convex hull $\conv(\Gamma^{\mathrm{A}\prec\mathrm{B}},\Gamma^{\mathrm{B}\prec\mathrm{A}})$ contains the vectors $\bm{h}_{\textrm{conv}} = (H(T|Q))_{T\in\mathcal{S}^{\rm c}}$ for causal correlations. The system of inequalities $I\bm{h}\le\bm{0}$ characterizing $\conv(\Gamma^{\mathrm{A}\prec\mathrm{B}},\Gamma^{\mathrm{B}\prec\mathrm{A}})$ can then again be transformed in a similar way to Eq.~\eqref{eq:lin_tran} into a new system $\tilde{I}\tilde{\bm{h}}\le\bm{0}$ defining the cone of corresponding entropy vectors $\tilde{\bm{h}}=(H(T))_{T\in\widetilde{\mathcal{S}}^{\rm c}}$, which we again denote by $\conv_Q(\Gamma^{\mathrm{A}\prec\mathrm{B}},\Gamma^{\mathrm{B}\prec\mathrm{A}})$.
In contrast to the DAG-based method, the only constraint on $Q$ is, now, $H(Q)\le 1$, since $Q$ need not be independent of the counterfactual output variables $A_{xy},B_{xy}$.
Finally, we need to project onto the marginal scenario $\mathcal{M}$ in Eq.~\eqref{eq:counterfactualM}.
The causal polyhedron is thus given, in analogy to Eqs.~\eqref{eq:bip_cau_coneQ} and~\eqref{eq:bip_cau_cone}, by
\begin{align}
	\label{eq:cone_counter_bi}
	\Gamma^{\mathrm{causal}}_{\rm AB} =  \Pi_\mathcal{M} \Big[&\Gamma^{\widetilde{\mathcal{S}}}\cap \conv_Q(\Gamma^{\mathrm{A}\prec\mathrm{B}},\Gamma^{\mathrm{B}\prec\mathrm{A}}) \notag \\[-1mm]
& \hspace{14mm} \cap\textrm{L}_{\mathcal{C}}\big(\{H(Q)\leq 1\}\big) \Big],
\end{align}
where we have $\Gamma^{\widetilde{\mathcal{S}}}=\bigcap_{x,y\in\{0,1\}}\Gamma_{\{A_{xy},B_{xy},Q\}}$.

\subsubsection{Entropic causal inequalities for counterfactual variables and their violation}

As in Sec.~\ref{sec:causal_str_bi}, the construction above allows one to obtain the full list of entropic inequalities characterizing $\Gamma^{\mathrm{causal}}_{\rm AB}$. 
After removing the trivial inequalities directly implied by Shannon constraints on $\mathcal{M}$, we find that there are 6 nontrivial entropic causal inequalities, which can be grouped into two equivalence classes of inequalities under the relabeling of inputs:
\begin{equation}\label{eq:marg_ineq1}
	I(A_{00}:B_{00}) - H(A_{01}) - H(B_{10})\leq 1
\end{equation}
and 
\begin{align}\label{eq:marg_ineq2}
	& I(A_{00}:B_{00}) + I(A_{11}:B_{11})\notag\\
	& \quad - H(A_{01}B_{01}) - H(A_{10}B_{10})\leq 2.
\end{align}
The fact that these inequalities have nontrivial bounds is, as for the DAG-based method, a result of the constraint $H(Q)\le 1$ which means $\Gamma^{\mathrm{causal}}_{\rm AB}$ is a polyhedron characterized by a set of inhomogeneous inequalities.
Indeed, if one chooses not to eliminate $Q$ from the entropic description, one obtains a convex cone characterized by the above equations, except that the right-hand sides are multiplied by $H(Q)$ (see the discussion in Appendix~\ref{apndx:DAGineqs}).

In contrast to the case for the DAG-based approach, where violation of the causal inequalities we obtained was possible even with deterministic distributions, it is clear that such distributions provide no interesting behavior in the counterfactual approach since any such distribution will have a null entropy vector.
By looking at equal mixtures of deterministic causal distributions $P^{\mathrm{A}\prec \mathrm{B}}$ and $P^{\mathrm{B}\prec \mathrm{A}}$, however, we were able to verify that the inequalities in Eqs.~\eqref{eq:marg_ineq1}--\eqref{eq:marg_ineq2} can indeed be saturated by such (causal) distributions and are thus tight.

In order to study the potential violation of these entropic inequalities, we again need to look at nondeterministic distributions.
One can easily see, however, that Eqs.~\eqref{eq:marg_ineq1}--\eqref{eq:marg_ineq2} cannot be violated when restricted to distributions satisfying $H(A_{xy})\le 1$ and $H(B_{xy})\le 1$ for all $x,y\in\{0,1\}$, as this also implies that $I(A_{xy}:B_{xy})\le 1$.
This means that the inequalities for counterfactual variables are unable to detect noncausality when both parties are restricted to binary outputs.

To find violations we again look at the extremal rays of the Shannon cone $\Gamma^{\mathcal{S}}$ of Eq.~\eqref{eq:shannonCone_counterfact} which violate one of the inequalities, and examine whether these rays can be reached by any probability distribution.
Considering bounds on $H(A_{xy})$ and $H(B_{xy})$ strictly larger than $1$, we find that violations are possible for any such bound.
Moreover, the entropy vectors giving maximal violation of Eqs.~\eqref{eq:marg_ineq1} and~\eqref{eq:marg_ineq2} are generally realizable with equal mixtures of causal and noncausal distributions.
For example, given the constraints $H(A_{xy})\le \log_2 k$ and $H(B_{xy})\le \log_2 k$ for some integer $k \ge 2$, the distribution
\begin{equation}\label{eq:margScenViolatingDist}
	P_k(ab|xy)=\delta_{x,y}\ \frac{1}{k}\delta_{a,b}+(1{-}\delta_{x,y})\ \delta_{a,0}\delta_{b,0},
\end{equation}
where $a,b\in\{0,\dots,k-1\}$, realizes such an extremal point for all $k \ge 2$, and provides a violation of both Eqs.~\eqref{eq:marg_ineq1} and~\eqref{eq:marg_ineq2} for $k > 2$.
For $k=2$ (binary outputs), this distribution can be written as the convex combination
\begin{equation}
	P_2(ab|xy)=\frac{1}{2} P^{\textrm{NC}}(ab|xy) + \frac{1}{2} P^{\textrm{C}}(ab|xy),
\end{equation}
where $P^{\textrm{NC}}(ab|xy)=\delta_{a\oplus x \oplus 1,y}\delta_{b\oplus y\oplus 1,x}$ maximally violates a GYNI-type inequality (it is simply a symmetry of the third distribution in Eq.~\eqref{eq:GYNIdists}, obtained by flipping all outputs), and $P^{\textrm{C}}(ab|xy)=\delta_{a,0}\delta_{b,0}$ is causal.
Even though it does not violate Eq.~\eqref{eq:marg_ineq1} or~\eqref{eq:marg_ineq2}, $P_2$ is noncausal.
The distribution $P_k$ can be seen as a possible generalization of a GYNI-violating distribution.

This link to the GYNI-type inequalities and correlations can be made more explicit by considering the related distribution
\begin{equation}\label{eq:GYNIdistGen}
	P_k'(ab|xy)=\delta_{x,y}\ \frac{1}{k{-}1}\delta_{a,b}(1{-}\delta_{a,0}\delta_{b,0}) +  (1{-}\delta_{x,y})\ \delta_{a,0}\delta_{b,0},
\end{equation}
with again $a,b\in\{0,\dots,k-1\}$.
We have $P_2'=P^{\textrm{NC}}$, and, for $k \ge 3$, $P_k'$ has the same entropy vector as $P_{k-1}$.
$P_k'$ can be clearly simulated from $P_2'=P^{\textrm{NC}}$ by making use of shared randomness and by letting both parties replace the output $1$ obtained from $P_2'$ by a shared random value $a=b \in \{1,\dots,k-1\}$. 
It is interesting to see, then, that the GYNI-maximally-violating distributions also provide the best behavior entropically when augmented with shared randomness, even though they fail to violate the inequalities when the parties have only binary outputs.

As for the DAG-based method, it is also interesting to look at the sensitivity of the inequalities with respect to the detection of noncausality.
To do so, we again looked at distributions of the form given in Eq.~\eqref{eq:mix_distr}, but where $P^{\mathrm{NC}}$ and $P^{\mathrm{C}}$ are now equal mixtures of 3-outcome deterministic noncausal and causal distributions, respectively.
The number of such distributions makes an exhaustive search difficult, but by sampling randomly we were nonetheless able to verify the existence of distributions $P^{\varepsilon}(ab|xy)$ which violate the entropic inequalities~\eqref{eq:marg_ineq1} and~\eqref{eq:marg_ineq2} for arbitrary small $\varepsilon$.
Although the examples we found are not particularly informative or simple (and thus we refrain from giving them explicitly), they nevertheless show that the entropic inequalities~\eqref{eq:marg_ineq1} and~\eqref{eq:marg_ineq2} exhibit the desired sensitivity.

Finally, one may again ask whether one can violate any of the entropic inequalities for counterfactuals within the process matrix formalism, or whether any noncausal correlation can be mixed with a causal one to violate an entropic inequality, as is the case for entropic Bell inequalities obtained from the counterfactual approach~\cite{Chaves:2013aa}.
We leave this as an open question, but note only that we were not able to find a way to do so: for example, we were unable to find a violation (with or without the use of shared randomness) for noncausal distributions realizable within the process matrix framework.

\section{Multipartite entropic causal inequalities}
\label{sec:multipartite}

The notion of causal correlations can be extended to more than two parties in a recursive manner~\cite{Abbott:2016aa,oreshkov15}.
Consider $N$ parties ${\rm A}_1,\dots,{\rm A}_N$, with inputs $\bm{x}=(x_1,\dots,x_N)$ and outputs $\bm{a}=(a_1,\dots,a_N)$.
In any given run, one party, say ${\rm A}_k$, must act first, and none of the other parties can signal to them, which implies $P(a_k|\bm{x})=P(a_k|x_k)$.
The correlations shared by the remaining $N-1$ parties, conditioned on the input and output of the first, must also in turn be causal.
However, note that the causal order itself (and not only the response functions) of the remaining parties may depend on the input and output of the first, a phenomenon called dynamical causal order~\cite{hardy2005probability,oreshkov15,Abbott:2016aa}, and which goes beyond the standard model of fixed causal Bayesian networks.

An $N$-partite correlation $P(\bm{a}|\bm{x})$ is thus called causal if it can be decomposed in the following way~\cite{Abbott:2016aa,oreshkov15}:
\begin{equation}\label{eq:causal_corr}
	P(\bm{a}|\bm{x})=\sum_{k=1}^N q_k\, P_k(a_k|x_k) \, P_{k,x_k,a_k}(\bm{a}_{\backslash k}|\bm{x}_{\backslash k}),
\end{equation}
where $\bm{x}_{\backslash k}=(x_1,\dots,x_{k-1},x_{k+1},\dots,x_N)$ and $\bm{a}_{\backslash k}=(a_1,\dots,a_{k-1},a_{k+1},\dots,a_N)$, with $q_k \ge 0$, $\sum_k q_k = 1$, and where for each $k, x_k, a_k$, $P_{k,x_k,a_k}(\bm{a}_{\backslash k}|\bm{x}_{\backslash k})$ is a causal $(N{-}1)$-partite correlation (down to the lowest level of this recursive definition, where any $1$-partite correlation is considered to be causal).
Note that, for $N=2$ this reduces to Eq.~\eqref{eq:causalCorreltn}.
The entropic approach can be generalized to the multipartite scenario using a similar recursive method.

\subsection{Causal Bayesian network method}

\begin{figure}[t]
	\begin{center}
		\includegraphics[width=.25\textwidth]{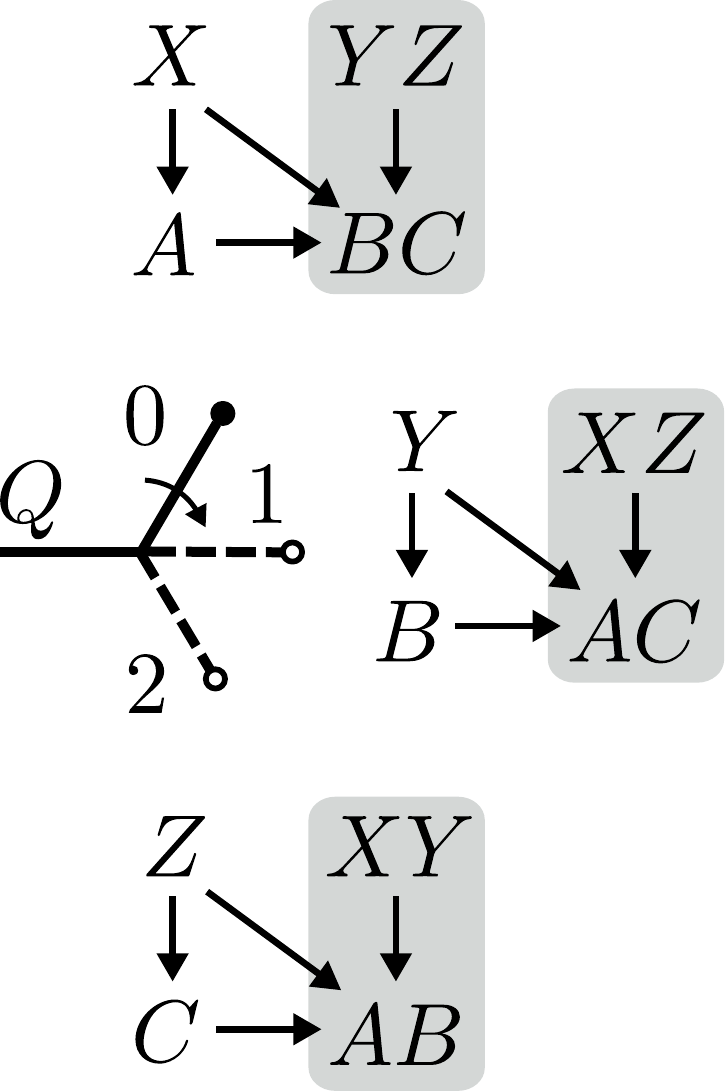}
	\end{center}
	\caption{DAGs for tripartite causal correlations. The latent ``switch'' variable $Q$ determines which DAG is ``activated''. Correlations among variables from shaded rectangles are causal conditionally on the input and output of the party acting first.  \label{fig:dag_tri}}
\end{figure}

It is instructive to first look into the details of the tripartite case -- in which case we shall denote the parties Alice (${\rm A}$), Bob (${\rm B}$) and Charlie (${\rm C}$), as is standard -- before generalizing the method to more parties.
The general method follows that used for the bipartite case in Sec.~\ref{sec:causal_str_bi}, and the relevant conditional DAG is shown in Fig.~\ref{fig:dag_tri}.
The set of observable variables to be considered here is $S = \{X,Y,Z,A,B,C\}$, the marginal scenario and the probability structure are ${\mathcal M} = {\mathcal S} = \{S\}$.

The polytope of tripartite causal correlations (i.e., of the form Eq.~\eqref{eq:causal_corr}) can be written as
\begin{equation}
	\mathcal{P}_{\rm ABC}^{\rm causal} = \conv(\mathcal{P}^{\rm A},\mathcal{P}^{\rm B},\mathcal{P}^{\rm C}),
\end{equation}
where $\mathcal{P}^{\rm A}$ is the polytope of causal distributions consistent with Alice acting first and such that the remaining conditional correlation shared by Bob and Charlie is causal, and analogously for $\mathcal{P}^{\rm B}$ and $\mathcal{P}^{\rm C}$.
As a consequence, in order to define the polyhedron characterizing entropically tripartite causal correlations, which we denote $\Gamma^{\rm causal}_{\rm ABC}$, we first need to define the corresponding polyhedra, namely $\Gamma^{\rm A},\Gamma^{\rm B},\Gamma^{\rm C}$, associated with each party acting first.

Let us thus consider $\Gamma^{\rm A}$. 
According to the recursive definition given in Eq.~\eqref{eq:causal_corr}, for any $x,a$, the conditional entropy vector $\bm{h}^{xa}_{\rm BC}=(H(T|X=x,A=a))_{T\subset \{Y,Z,B,C\}}$ for a correlation in $\mathcal{P}^{\rm A}$ must be contained in the bipartite causal polyhedron $\Gamma^{\rm causal}_{\rm BC}$, defined for Bob and Charlie as in Eqs.~\eqref{eq:bip_cau_coneQ}--\eqref{eq:bip_cau_cone}. 
By convexity this also implies that $\bm{h}_{\rm BC}=(H(T|XA))_{T\subset \{Y,Z,B,C\}}= \sum_{x,a} P(x,a) \bm{h}^{xa}_{\rm BC}$ is in $\Gamma^{\rm causal}_{\rm BC}$. 
We can then use a similar transformation to Eq.~\eqref{eq:lin_tran} to obtain constraints on $\Gamma^{\rm A}$: if entropy vectors $\bm{h}_{\rm BC}$ in $\Gamma^{\rm causal}_{\rm BC}$ satisfy the inequalities $\bm{I}\cdot\bm{h}_{\rm BC}\le \beta$, then the corresponding (unconditional) entropy vector $\bm{h}=(H(T))_{T\subset S}$ must satisfy the inequalities $\mathcal{T}_{XA}(\bm{I})\cdot\bm{h}\le\beta$.
Writing $\mathcal{T}^*_{XA}$ for the dual transformation on the space of entropy vectors, we thus have that $\bm{h}\in\mathcal{T}^*_{XA}(\Gamma_{\rm BC}^{\textrm{causal}})$. 
Together with the facts that $\bm{h}$ must lie in the Shannon cone $\Gamma_S$ for the relevant variables, that all the inputs must be independent from each other, and that Alice's output must be independent from Bob and Charlie's inputs (conditioned on her input), we obtain the characterization
\begin{align}\label{eq:con_A_first}
\Gamma^{\rm A} =\,&  \Gamma_S \cap \mathcal{T}^*_{XA}(\Gamma_{\rm BC}^{\rm causal})  \notag\\
& \quad \cap \textrm{L}_{\mathcal{C}}(\{ X \CI Y \CI Z, A \CI YZ | X \}),
\end{align}
with similar expressions for $\Gamma^{\rm B}$ and $\Gamma^{\rm C}$.

Following the same approach as in Sec.~\ref{sec:causal_str_bi}, we introduce a (now three-valued) switch variable $Q$ (see Fig.~\ref{fig:dag_tri}). 
Similarly to what we observed in the bipartite case, the convex hull $\conv(\Gamma^{\rm A},\Gamma^{\rm B},\Gamma^{\rm C})$ contains the conditional entropy vectors $(H(T|Q))_{T\subset S}$ for tripartite causal correlations. 
The inequalities characterizing $\conv(\Gamma^{\rm A},\Gamma^{\rm B},\Gamma^{\rm C})$ can again be transformed into inequalities satisfied by the entropy vector $\tilde{\bm{h}}= (H(T))_{T\subset \tilde S}$, for variables in $\tilde S = S\cup\{Q\}$, by introducing a transformation $\mathcal{T}_Q$ as in Eq.~\eqref{eq:lin_tran}, thus defining the polyhedron ${\rm conv}_{Q}(\Gamma^{\rm A},\Gamma^{\rm B},\Gamma^{\rm C} )$ as before. 
Taking into account the Shannon constraints for all variables in $\tilde S$, the independence constraints ${\rm CI}_{Q} = (X\CI Y \CI Z \CI Q)$ and the bound $H(Q)\leq \log_2 3$, and finally projecting onto the observable variables in $S$, we see that the entropy vectors for tripartite causal correlations belong to the polyhedron
\begin{align}
\hspace{-2mm} (\Gamma^{\mathrm{causal}}_{\rm ABC})_0 = & \ \Pi_S \left[ \Gamma_{\tilde S} \cap {\rm conv}_{Q}(\Gamma^{\rm A},\Gamma^{\rm B},\Gamma^{\rm C} )  \right.\notag\\[-1mm]
  & \qquad \quad \cap \textrm{L}_{\mathcal{C}}(\{{\rm CI}_{Q}, \, H(Q)\leq \log_2 3\}\big)\Big].
\end{align}

While this characterization is certainly valid, some subtleties arising from the differences between the probabilistic and entropic descriptions allow one to actually make it tighter.
Specifically, certain conditions implied by the definition~\eqref{eq:causal_corr} need not be implied by the corresponding entropic definition outlined above.
For example, if $P(abc|xyz)$ is a causal correlation, then the bipartite marginal distributions $P_{x}(bc|yz)=\sum_{a}P(abc|xyz)$ and $P(bc|yz)=\sum_{x}P(x)P_{x}(bc|yz)$ are both causal (as are the corresponding marginals for each other pair of parties)~\cite{Abbott:2016aa}.
This implies that the entropy vectors $(H(T|X))_{T\subset\{Y,Z,B,C\}}$ and $(H(T))_{T\subset\{Y,Z,B,C\}}$ corresponding to a tripartite causal correlation must also satisfy all the inequalities characterizing the bipartite causal polyhedron $\Gamma^{\rm causal}_{\rm BC}$ -- which may not necessarily be implied by the characterization of $(\Gamma^{\mathrm{causal}}_{\rm ABC})_0$ above.
We can thus tighten the previous characterization, and define the tripartite causal polyhedron as\footnote{In Eq.~\eqref{eq:tri_causal_polyhedron} we abuse the notation slightly and denote by $\Gamma_{\rm BC}^{\rm causal}$ the set of entropy vectors $(H(T))_{T\subset S}$ -- instead of $(H(T))_{T\subset \{Y,Z,B,C\}}$ -- which satisfy the constraints characterizing $\Gamma_{\rm BC}^{\rm causal}$ as defined in Eqs.~\eqref{eq:bip_cau_coneQ}--\eqref{eq:bip_cau_cone}. The transformation $\mathcal{T}_{X}$, of which $\mathcal{T}^*_{X}$ is the dual, is again defined in a similar way as in Eq.~\eqref{eq:lin_tran}.}
\begin{equation} \label{eq:tri_causal_polyhedron}
	\Gamma_{\rm ABC}^{\rm causal} = (\Gamma^{\mathrm{causal}}_{\rm ABC})_0 \cap \Gamma_{\rm BC}^{\rm causal} \cap \mathcal{T}^*_{X}(\Gamma_{\rm BC}^{\rm causal}) \cap \text{[perms.]},
\end{equation}
where $\text{[perms.]}$ denotes the permutations of the preceding two terms for the other parties.
Note that such extra constraints do not need to be imposed in the bipartite case since the causality of all one-party marginals is equivalent to them being valid probability distributions, which is already assured by the elemental inequalities.

To extend the above idea to the general multipartite case of Eq.~\eqref{eq:causal_corr}, we simply define recursively (here the notation should be self-evident)
\begin{align}
	\Gamma^{{\rm A}_k} =  \Gamma_{\{\bm{X},\bm{A}\}}\cap \mathcal{T}^*_{X_k A_k}\left(\Gamma_{\bm{\mathrm{A}}_{\backslash k}}^{\rm causal}\right) \cap  \textrm{L}_{\mathcal{C}}({\rm CI}_{{\rm A}_k}),  
\end{align}
where ${\rm CI}_{{\rm A}_k}$ denotes the set of independence constraints resulting from the assumption that all parties' inputs are independent, i.e.~$X_1 \CI \dots \CI X_N$, and that party $k$ acts first, which implies $A_k \CI \bm{X}_{\backslash k}|X_k$.
The causal polyhedron is then defined as
\begin{align}\label{eq:con_cau_mult}
\Gamma^{\rm causal}_{\bm{\mathrm{A}}}= & \ \Pi_{\bm{X},\bm{A}} \left[ \Gamma_{\{\bm{X},\bm{A},Q\}}  \cap {\rm conv}_{Q}(\{\Gamma^{{\rm A}_k}\}_k )   \right. \notag\\[-1mm]
& \hspace{11mm} \cap \textrm{L}_{\mathcal{C}}(\{{\rm CI}_{Q}, \, H(Q)\leq \log_2 N\}\big)\Big]\notag\\
& \ \bigcap_k \left[ \Gamma_{\bm{\mathrm{A}}_{\backslash k}}^{\rm causal} \cap 	\mathcal{T}^*_{X_k}\left(\Gamma_{\bm{\mathrm{A}}_{\backslash k}}^{\rm causal}\right) \right],
\end{align}
where ${\rm CI}_{Q}$ denotes the independence relation between all inputs and $Q$, i.e.~$X_1 \CI \cdots \CI X_N \CI Q$.

\subsection{Counterfactual variable method}\label{sec:count_mult}

A similar generalization is possible also for the counterfactual method. 
Again, it is instructive to look first at the tripartite case, where the set of variables to be considered is $S = \{A_{xyz},B_{xyz},C_{xyz}\}_{x,y,z}$, the marginal scenario is ${\mathcal M} = \{\{A_{xyz},B_{xyz},C_{xyz}\}\}_{x,y,z}$ and we take the probability structure to be ${\mathcal S} = {\mathcal M}$.
We start by defining the polyhedron for the case in which Alice acts first,
\begin{align}
\Gamma^{\rm A} =  \bigcap_{xyz} \Big[& \Gamma_{\{A_{xyz},B_{xyz},C_{xyz}\}} \cap \mathcal{T}^*_{A_{xyz}}(\Gamma_{\rm BC}^{\rm causal})  \notag\\[-3mm]
& \qquad \cap \textrm{L}_{\mathcal{C}}\big( \{A_{xyz} = A_{xy'z'} \}_{y'z'}\big)\Big],
\end{align}
which is the analogue, for the counterfactual method, of the polyhedron in Eq.~\eqref{eq:con_A_first}. Similar definitions hold for $\Gamma^{\rm B}$ and $\Gamma^{\rm C}$. 
The tripartite polyhedron of causal counterfactual inequalities can then be defined, following a similar reasoning to the previous case, as
\begin{align}\label{eq:count_trip_pol}
&\Gamma_{\rm ABC}^{\rm causal} = \Pi_{\mathcal{M}}\! \Big[ \Gamma^{\widetilde{\mathcal{S}}}\cap {\rm conv}_{Q}(\Gamma^{\rm A},\Gamma^{\rm B},\Gamma^{\rm C} ) \notag\\[-2mm]
& \hspace{30mm} \cap \textrm{L}_{\mathcal{C}}\!\big(\!\{\!H(Q)\!\leq\! \log_2 \!3\}\!\big)\!\Big]\notag\\
& \hspace{15mm} \bigcap_x \Gamma_{{\rm BC}|x}^{\rm causal} \ \bigcap_y \Gamma_{{\rm AC}|y}^{\rm causal} \ \bigcap_z \Gamma_{{\rm AB}|z}^{\rm causal},
\end{align}
where $\widetilde{\mathcal{S}}=\{ \{A_{xyz},B_{xyz},C_{xyz},Q\}\}_{x,y,z}$ and
$\Gamma_{{\rm BC}|x}^{\rm causal}$ is defined by imposing the constraints characterizing $\Gamma_{\rm BC}$ (\emph{a priori} defined for some variables $B_{yz}, C_{yz}$) to the variables $B_{xyz}, C_{xyz}$, and with similar definitions for $\Gamma_{{\rm AC}|y}^{\rm causal}$ and $\Gamma_{{\rm AB}|z}^{\rm causal}$.

As for the case based on causal Bayesian networks, the construction in Eq.~\eqref{eq:count_trip_pol} can then be generalized to an arbitrary number of parties in a recursive way.

\section{Information bounds in causal games}\label{sec:infoBounds}

One of the advantages of the entropic approach is that it allows information theoretic constraints to be naturally imposed, derived, and interpreted~\cite{Pawlowski2009,Chaves2015}.
As an illustration, we consider a simple application of our approach to understanding the role of bounded communication in causal games.

Consider the generalization of the GYNI game described in Sec.~\ref{sec:causalcorr} to arbitrary numbers of inputs and outputs, in which two parties try to maximize the winning probability $p_{\mathrm{succ}}=P(a=y,b=x)$.
If the parties operate causally, then in any given round of the game only one-way communication may occur.
One may be interested in the effect of limiting the amount communication that can occur in any such round.
In the entropic framework, this can easily be taken into account by adding an additional constraint of the form $I(X:B)\leq I_{\mathrm{max}}$ to $\Gamma^{\mathrm{A}\prec\mathrm{B}}$ in order to restrict $B$'s dependency on $X$, and similarly imposing  $I(Y:A)\leq I_{\mathrm{max}}$ to  $\Gamma^{\mathrm{B}\prec\mathrm{A}}$, where the quantity $I_{\mathrm{max}}$ represents the maximum allowable entropy of the classical message communicated by the parties.
For example, if the parties are permitted, in each round, to exchange a classical $d$-dimensional system, then $I_{\mathrm{max}}=\log_2 d$.
In general, the amount of one-way communication $I_{\mathrm{max}}$ does not need to be specified in advance, it will appear as a parameter in our inequalities. 
By applying the approach of Sec.~\ref{sec:causal_str_bi} to this scenario one finds that causal correlations must then obey the inequality
\begin{equation}\label{eq:inf_caus_simpler}
I(X:B) + I(Y:A) \leq I_{\mathrm{max}},
\end{equation}
i.e., the sum of the two mutual informations is similarly bounded by $I_{\mathrm{max}}$.
Although this is perhaps not unexpected, it shows the ease with which such bounds can be derived in the entropic framework.

A more subtle variant is obtained by considering a slight generalization of the causal game proposed by Oreshkov, Costa, and Brukner (OCB) in Ref.~\cite{oreshkov12}.
In this game, the goal is also for one party to guess the other party's input; in contrast to the GYNI game, however, an additional input random bit $Y'$ is given,\footnote{In the original OCB game, only Bob receives the input $Y'$, whereas in the variant we consider here, both parties have access to it.} which determines whether it is Bob who should guess Alice's input (if $Y'=0$) or vice versa (if $Y'=1$).
The parties thus now attempt to maximize the winning probability 
\begin{equation}
p_{\mathrm{succ}} = \frac{1}{2}\Big(P(b=x\,|\,Y'=0) + P(a=y\,|\,Y'=1)\Big).
\end{equation}
An analogous entropic inequality can be obtained via a combination of the methods discussed in Sec.~\ref{sec:bipartite}. 
Since the relevant direction of communication in each round of this game depends on the additional input $Y'$, we will combine the DAG-based method for the variables $X,Y,A,B$ with the counterfactual approach to condition on $Y'$. 
More precisely, one may take $\mathcal{S}=\mathcal{M} =  \big\{ \{X,Y,A_{y'},B_{y'}\} \big\}_{y'}$ and  $\widetilde{\mathcal{S}} = \big\{ \{X,Y,A_{y'},B_{y'},Q\} \big\}_{y'}$;
the relevant causal constraints for the cones $\Gamma^{\mathrm{A}\prec\mathrm{B}}$ and $\Gamma^{\mathrm{B}\prec\mathrm{A}}$ and the polyhedron $\widetilde{\Gamma}^{\text{causal}}_{\rm AB}$ are the same as those imposed on $X,Y,A,B,Q$ in the DAG-based method, except that now they are applied to each copy of the conditional variables $A_{y'}$ and $B_{y'}$, and the communication bounds $I(X:B_{y'})\leq I_{\mathrm{max}}$ and $I(Y:A_{y'})\leq I_{\mathrm{max}}$ are imposed on the corresponding cones.
Notice that, in this way, we are assuming that  $X \CI Y \CI Y' \CI Q$. 
Combining the above constraints with the analysis in Sec.~\ref{sec:bipartite}, one finds that causal correlations must obey
\begin{equation}
\label{eq:inf_caus}
I(X:B\,|\,Y'=0) + I(Y:A\,|\,Y'=1) \leq I_{\mathrm{max}}.
\end{equation}

This inequality, for the special case of binary inputs and outputs and with $I_{\mathrm{max}}=1$, was proposed in 
Ref.~\cite{Ibnouhsein2015} as a potential principle to bound the set of correlations obtainable within the process matrix formalism,\footnote{Ref.~\cite{Ibnouhsein2015} proposed this inequality in the framework of the original OCB game. However, one can easily see that our derivation of Eq.~\eqref{eq:inf_caus} in the more general scenario implies that it must hold in that framework too. Indeed, if only Bob receives $Y'$, then this implies the additional constraint $H(A_{0})=H(A_1)$ when $\mathrm{A}\prec\mathrm{B}$. The set of correlations obtainable is thus a subset of those obtainable in the more general version of the game, and thus Eq.~\eqref{eq:inf_caus} must again hold true.} in analogy with the celebrated information causality principle~\cite{Pawlowski2009} that provides bounds on the strength of bipartite quantum correlations. 
Our approach allowed us to show that Eq.~\eqref{eq:inf_caus} indeed holds for causal processes, but it remains to be seen whether such a constraint on mutual informations for causal correlations can be violated within the process matrix framework. 
This example, however, highlights the potential of the entropic approach to causal correlations for studying information-theoretic principles.

\section{Discussion}

Since Bell first formulated his eponymous theorem, understanding the role of causality within quantum mechanics has been a central yet thorny goal. 
Complicating matters further, the very idea of a definite causal order itself has begun to be questioned.
While sophisticated frameworks have been introduced in an effort to free physical theories from the shackles of a rigid causal framework, the issue of whether nature permits violations of causal inequalities remains an elusive question.

Against this backdrop, our aim in this paper was to introduce an entropic approach to studying causal correlations, and to this end we presented two complementary methods: the first based on the consideration of the entropies of the variables appearing in the causal Bayesian networks describing causal scenarios, and the second based on a counterfactual description of the outcome variables appearing in such networks. 
Focusing on bipartite causal scenarios, we described in detail the successful application of both methods to derive nontrivial entropic causal inequalities, before showing how the characterizations can be generalized to multipartite scenarios.
In contrast to the usual approach to causal correlations based on probability distributions, the entropic causal inequalities we derived using both methods are valid for any finite number of possible outcomes, as well as for any number of inputs for the first method based on causal Bayesian networks, and thus provide a very concise description of causal correlations. 
We discussed the ability for the derived entropic causal inequalities to witness the noncausality of several classes of interesting noncausal correlations, but were nonetheless unable to find violations of the inequalities by correlations obtainable within the process matrix formalism~\cite{oreshkov12} using qubit systems.
In light of the coarse-grained description provided by entropic inequalities and the fact that the known violations of standard causal inequalities are in general rather small~\cite{Branciard:2016aa}, this is arguably an unsurprising negative result. 
The question of whether entropic causal inequalities can be violated within the process matrix formalism and (more importantly) by quantum correlations thus remains open.
More generally, our construction can be used to characterize arbitrary convex combinations of different causal Bayesian networks, and thus provides, for example, a natural tool to investigate stronger notions of multipartite Bell nonlocality~\cite{Svetlichny1987,Gallego2012,Bancal2013,Chaves2016causal} from the entropic perspective. 

In view of this new framework for the study of causal correlations we believe that several other directions of research can naturally be pursued. 
Here we focused on using the Shannon entropies of the relevant variables, but it is known that, at least in particular scenarios, the same approach can be used to derive constraints using certain generalized entropies~\cite{Rastegin2014,Wajs2015} and even with non-statistical information measures such as the Kolmogorov complexity~\cite{Chaves2014b}. 
Can our framework be extended to these other information measures, and if so, are they more sensitive to violations of causality? 
Similarly, one may wonder whether the addition of non-Shannon-type inequalities to the entropic descriptions of causal correlations considered might lead to tighter constraints~\cite{zhang98,Fritz:2013aa,weilenmann16}.
Moreover, in the multipartite characterization of causal correlations in Sec.~\ref{sec:multipartite} we also saw that causality imposes additional entropic constraints on marginal and conditional distributions that allow us to tighten our characterization in Eq.~\eqref{eq:tri_causal_polyhedron}. 
It remains an open question whether additional such constraints can be found that tighten even further the characterization.

Another important direction to consider would be the ability to formulate, and perhaps violate, information-theoretical principles~\cite{Chaves2015} of causality.
We provided, as a simple application, an idea for one possible approach, showing how simple bounds on mutual informations can be derived for causal games where communication is limited in each direction.
It would be interesting to see, in particular, whether such principles could be violated within the process matrix formalism and, if so, the connection to the violation of causal inequalities. 
For example, does the violation of causal inequalities imply the violation of some principle implied by quantum mechanics?  
We expect our results to motivate these and many more future investigations.

\section*{Acknowledgments}

We acknowledge fruitful discussions with Philippe Allard
Gu\'erin, Flavio Baccari, and \v{C}aslav Brukner.
This work was funded by the DAAD, the ``Retour Post-Doctorants'' program (ANR-13-PDOC-0026) of the French National Research Agency, the Brazilian ministries MEC and MCTIC, and the FWF (Project: M 2107 Meitner-Programm).


\bibliography{entr_causal_ineqs}


\onecolumngrid
\appendix

\section{Causal correlations not contained in $\conv(\Gamma^{\mathrm{A}\prec\mathrm{B}},\Gamma^{\mathrm{B}\prec\mathrm{A}})$}
\label{apndx:counterex}

Starting with the systems of inequalities $I_0\bm{h}\le\bm{0}$ and $I_1\bm{h}\le\bm{0}$ characterizing the cones $\Gamma^{\mathrm{A}\prec\mathrm{B}}$ and $\Gamma^{\mathrm{B}\prec\mathrm{A}}$ defined in Eqs.~\eqref{eq:coneAB} and~\eqref{eq:coneBA}, the characterization $I\bm{h}\le\bm{0}$ of $\conv(\Gamma^{\mathrm{A}\prec\mathrm{B}},\Gamma^{\mathrm{B}\prec\mathrm{A}})$ can be found by first solving the extremal ray enumeration problem for the extremal rays of $\Gamma^{\mathrm{A}\prec\mathrm{B}}$ and $\Gamma^{\mathrm{B}\prec\mathrm{A}}$, taking the union of these rays and finally solving the facet enumeration problem for the inequalities characterizing $\conv(\Gamma^{\mathrm{A}\prec\mathrm{B}},\Gamma^{\mathrm{B}\prec\mathrm{A}})$.

We find that there are six nontrivial inequalities (i.e., non Shannon-type inequalities) for $\conv(\Gamma^{\mathrm{A}\prec\mathrm{B}},\Gamma^{\mathrm{B}\prec\mathrm{A}})$, which correspond to four equivalence classes of inequalities under exchange of parties:\footnote{For compactness we generically write entropic inequalities not just in terms of Shannon entropies (as defined in Eq.~\eqref{ShannonE}), but also in terms of conditional entropies (of the form $H(A|B) \coloneqq H(AB) - H(B)$), of mutual information ($I(A:B) \coloneqq H(A)+H(B) - H(AB)$) and of conditional mutual information ($I(A:B|C)\coloneqq H(AC)+H(BC)-H(ABC)-H(C)$). The expressions given for the inequalities are of course not unique.}
\begin{align}
\label{eq:dag_conv_hull}
& I(X:YA) + I(Y:XB) - I(XY:AB) \leq 0\notag \\
& I(A:B) - I(A:B|X) - I(A:B|Y) \leq 0\notag\\
& I(X:A|B) - I(XB:A|Y) \leq 0\\
& I(A:B|X) - I(A:B|XY) - I(Y:B) \leq 0.\notag
\end{align}

In order to see that there are causal bipartite correlations that have entropy vectors not contained in $\conv(\Gamma^{\mathrm{A}\prec\mathrm{B}},\Gamma^{\mathrm{B}\prec\mathrm{A}})$, consider the following counterexample.
Take $P^{\mathrm{A}\prec\mathrm{B}}(ab|xy)=\delta_{a,x}\delta_{b,x}$ and $P^{\mathrm{B}\prec\mathrm{A}}(ab|xy)=\delta_{a,y}\delta_{b,y}$ and consider the inputs $x,y$ to be uniformly distributed so that $P^{\mathrm{A}\prec\mathrm{B}}(xyab)=\frac{1}{4}P^{\mathrm{A}\prec\mathrm{B}}(ab|xy)$ and $P^{\mathrm{B}\prec\mathrm{A}}(xyab)=\frac{1}{4}P^{\mathrm{B}\prec\mathrm{A}}(ab|xy)$.
The distribution $P(xyab)=\frac{1}{2}P^{\mathrm{A}\prec\mathrm{B}}(xyab)+\frac{1}{2}P^{\mathrm{B}\prec\mathrm{A}}(xyab)$ thus also defines a causal correlation $P(ab|xy)$, but one can verify that the entropy vector for $P(xyab)$ violates the first and last inequalities in~\eqref{eq:dag_conv_hull} with a value for the left-hand sides of $1-\frac{3}{2}\log_2\frac{3}{2}\approx 0.123 > 0$.

A similar conclusion can also be reached for the method based on counterfactual variables: starting from the definitions of $\Gamma^{\mathrm{A}\prec\mathrm{B}}$ and $\Gamma^{\mathrm{B}\prec\mathrm{A}}$ in Eqs.~\eqref{eq:coneABcoutnerfact} and~\eqref{eq:coneBAcoutnerfact} one finds that the inequalities characterizing $\conv(\Gamma^{\mathrm{A}\prec\mathrm{B}},\Gamma^{\mathrm{B}\prec\mathrm{A}})$ are precisely the same as the causal inequalities in Eqs.~\eqref{eq:marg_ineq1} and Eq.~\eqref{eq:marg_ineq2} except with bounds on the right-hand side of $0$.
One can easily verify that Eqs.~\eqref{eq:marg_ineq1} and Eq.~\eqref{eq:marg_ineq2} can be saturated by causal correlations (for some equal mixtures of correlations $P^{\mathrm{A}\prec \mathrm{B}}$ and $P^{\mathrm{B}\prec \mathrm{A}}$), thus providing such a counterexample.

\section{Bipartite entropic causal inequalities from the DAG method}
\label{apndx:DAGineqs}

The following is the full list of (equivalence classes of) entropic causal inequalities obtained from the DAG method, up to their symmetries under the exchange of parties.

Ten (of the twenty) families of inequalities have bounds of $0$ and can be violated by binary distributions:
\begin{align}
\label{eq:dag_v1}
& I(X:YA) + I(Y:XB) - H(AB) \leq 0\notag \\
& I(A:B) - I(A:B|X) - I(A:B|Y) -2H(AB|XY) \leq 0\notag\\
& I(X:YA) +I(Y:AB) -H(B|X) -H(A) \leq 0\notag\\
& I(A:B|X) - I(Y:B) -2H(B|XY) \leq 0\notag \\
& I(A:B|X) - I(A:B) - H(A|YB)-2H(B|XY)  \leq 0 \notag\\
& I(XA:Y) + I(YB:X) - H(X|YA) - H(A) \leq 0\\
& I(XA:Y) + I(YB:X) - H(B|YA) - H(A) \leq 0 \notag \\
& I(XA:Y) + I(YB:X) - I(A:B) -H(B|YA)-H(YA|X) \leq 0\notag\\
& I(XA:Y) + I(YB:X) - I(A:B) -H(B|YA) -H(AB|X) \leq 0 \notag \\
& I(XA:Y) + I(YB:X)- I(A:B) +I(X:A|Y) -H(XAB) \leq 0.\notag
\end{align}
Two more have non-zero bounds but, under the constraints that $H(A)\le 1$, $H(B)\le 1$, $H(X)\le 1$, $H(Y)\le 1$, turn out to be implied by the previous inequalities in Eq.~\eqref{eq:dag_v1}: 
\begin{align}
\label{eq:dag_v2}
&  H(X|B)+H(Y|A) - I(A:B|XY) -2H(X|YB) -2H(Y|XA) \le 1\notag\\
& I(XB:A) -3I(X:A) -3I(Y:B) - 4I(A:B|XY)-2H(XB|YA)+2H(B) \le 2.
\end{align}

Four correspond to ``corrected'' versions of the inequalities~\eqref{eq:dag_conv_hull} characterizing $\conv(\Gamma^{\mathrm{A}\prec\mathrm{B}},\Gamma^{\mathrm{B}\prec\mathrm{A}})$, and cannot be violated by binary distributions:
\begin{align}
\label{eq:dag_v3}
& I(X:YA) + I(Y:XB) - I(XY:AB) \le 1\notag \\
& I(A:B) - I(A:B|X) - I(A:B|Y) \le 2\notag\\
& I(X:A|B) - I(XB:A|Y) \le 1\\
& I(A:B|X) - I(A:B|XY) - I(Y:B) \le 1,\notag
\end{align}
while a further four can also not be violated by binary distributions:
\begin{align}
\label{eq:dag_v4}
& I(A:B|X) - I(A:B) - I(X:A|YB) - H(B|XY) \le 1\notag \\
& I(A:B|X) - I(A:B) - I(A:B|XY) - H(X|YB) \le 1\notag \\
& I(A:B|X) - I(A:B) - I(A:B|XY) - H(A|YB) \le 1 \\
& I(A:B|X) - I(A:B) - I(A:B|XY) + I(X:YA)+H(B|Y)-H(XB) \le 1\notag.
\end{align}

We note that, instead of projecting $\widetilde\Gamma^{\mathrm{causal}}_{\rm AB}$ (as defined in Eq.~\eqref{eq:bip_cau_coneQ}) onto the marginal scenario $\mathcal{M}=\big\{\{X,Y,A,B\}\big\}$ to obtain these entropic causal inequalities, one could start by projecting it onto the marginal scenario $\mathcal{M}'=\big\{\{X,Y,A,B\big\},\{Q\}\}$ which would amount to eliminating all entropies $H(T \cup \{Q\})$ for all nonempty subsets $T \subset \{X,Y,A,B\}$ from the description while keeping $H(Q)$.
By doing so, one obtains the same inequalities given in Eqs.~\eqref{eq:dag_v1} to~\eqref{eq:dag_v4}, except with all the right-hand sides multiplied by $H(Q)$.
The inequalities in Eq.~\eqref{eq:dag_v1} thus have no dependence on $H(Q)$ (i.e., on the exent to which correlations of different causal orders are mixed), while the remaining inequalities have a nontrivial dependence on it.
By eliminating $H(Q)$ using the constraint $H(Q)\le 1$ one then recovers the entropic causal inequalities above.\footnote{A similar procedure can also be followed for the approach with counterfactual variables, in which case one obtains upper-bounds of $H(Q)$ and $2H(Q)$ in Eqs.~\eqref{eq:marg_ineq1} and~\eqref{eq:marg_ineq2} (or $0$ for fixed-order correlations when $H(Q)=0$), before eliminating $H(Q)$ and obtaining Eqs.~\eqref{eq:marg_ineq1}--\eqref{eq:marg_ineq2} again.}

The inequalities containing $H(Q)$ may be of interest if, for some reason, one puts a nontrivial bound on $H(Q)$ (e.g., if one knows that one fixed causal order is more probable than the other), as they give novel constraints in such situations.
In the extreme case, if we know that $H(Q) = 0$, then the inequalities we obtain (namely Eqs.~\eqref{eq:dag_v1}--\eqref{eq:dag_v4}, with all upper bounds replaced by $0$) are valid for fixed-order causal correlations.
All of the inequalities in Eqs.~\eqref{eq:dag_v3}--\eqref{eq:dag_v4} with upper-bounds multiplied by $H(Q)$, except the second one in~\eqref{eq:dag_v3}, can be violated by binary noncausal correlations for any $H(Q)<1$; for the second inequality in Eq.~\eqref{eq:dag_v3} we were only able to find a violation for $H(Q)<\frac{1}{2}(1+\frac{3}{2}\log_2\frac{3}{2})\approx 0.939$.

\section{Relations between different probability structures}\label{sec:A_gp}

In the application of the counterfactual method to causal correlations discussed in Sec.~\ref{sec:counterfactuals_causal}, as a result of the structure of the marginal scenario one can prove that different choices of probability structure $\mathcal{S}$ give rise to the same observed marginal distributions.
This is due to the fact that since all the marginals $M_j\in\mathcal{M}$ are disjoint, they are always consistent with the global product probability distribution
\begin{equation}\label{eq:prodext}
P(a_{00},\ldots,b_{11}) = \prod_{xy} P(a_{xy},b_{xy}). 
\end{equation}
Hence, whichever probability structure $\mathcal{S}$ we choose (consistent with $\mathcal{M}$), the observed marginal probabilities can always be interpreted as arising from a global probability distribution.
Similarly, the choice of extended probability structure $\widetilde{\mathcal{S}}$ including the switch variable $Q$ in Eq.~\eqref{eq:tildeS} implies also the existence of a global probability distribution
\begin{equation}\label{eq:prodcondext}
	P(a_{00},\ldots,b_{11},q) = P(q)\prod_{xy} P(a_{xy},b_{xy}|q). 
\end{equation}
(Such a construction is also possible in some other types of scenarios; see Ref.~\cite{Vorob1962} for more general results.)
It thus follows that the probability structures $\widetilde{\mathcal{S}}$ that we chose and $\widetilde{\mathcal{S}}'=\{\tilde{S}\}$ again give rise to the same marginal distributions on $\mathcal{M}$.
A similar analysis can also be applied to the recursive method presented for the multipartite case in Sec.~\ref{sec:count_mult}.

At the level of entropic inequalities, however, the fact that we are considering Shannon inequalities that provide only an outer approximation of the entropy cone means that one may \emph{a priori} obtain different constraints depending on which of these equivalent probability structures one assumes. 
For the specific case of a marginal scenario with disjoint elements, i.e., $M_i\cap M_j = \emptyset$ for all $M_i,M_j\in \mathcal{M}$, a result by Mat\'u\v{s} (see Remark 1 in Ref.~\cite{matus2007}) implies, nevertheless, that choosing $\mathcal{S}=\mathcal{M}$ or $\mathcal{S}'=\{S\}$, with $S=\cup_{M_i\in \mathcal{M}} M_i$, also provides an equivalent description for the Shannon cone. More precisely, we have
\begin{equation}\label{eq:adh_disj}
	\Pi_{\mathcal{M}}( \Gamma^{\mathcal{S}'} ) = \Pi_{\mathcal{M}}( \Gamma_S ) = \Pi_{\mathcal{M}}\big( \Gamma_S \cap \textrm{L}_{\mathcal{C}}(\{\bm{X}_{M_i} \CI \bm{X}_{M_j}\}_{M_i,M_j\in \mathcal{M},i\neq j})\big) = \Pi_{\mathcal{M}} ( \bigcap_{M_i\in \mathcal{M}} \Gamma_{M_i} ) = \Pi_{\mathcal{M}}( \Gamma^{\mathcal{S}} ),
\end{equation}
where  $\bm{X}_{M_i}$ denotes the joint random variable associated with the subset of variables $M_i\in\mathcal{M}$. The linear constraints in Eqs.~\eqref{eq:coneABcoutnerfact} and~\eqref{eq:coneBAcoutnerfact} can then be imposed after the projection. Hence, the use of $\mathcal{S}=\mathcal{M}$ or $\mathcal{S}'=\{S\}$ is equivalent, in this case, even at the level of the Shannon cone description of $\Gamma^{\mathrm{A}\prec\mathrm{B}}$ and $\Gamma^{\mathrm{B}\prec\mathrm{A}}$.

One may hope that a similar analysis can be applied to show the equivalence of the probability structures $\widetilde{\mathcal{S}}$ and $\widetilde{\mathcal{S}}'=\{\tilde{S}\}$, where $\tilde{S}_i\cap \tilde{S}_j=\{Q\}$ for all distinct $\tilde{S}_i,\tilde{S}_j\in \widetilde{\mathcal{S}}$. 
However, even though the marginal scenario of interest remains the same as above, one no longer has $\widetilde{\mathcal{S}}=\mathcal{M}$ and, moreover, Eq.~\eqref{eq:cone_counter_bi} involves extra constraints given by $\conv_Q(\Gamma^{\mathrm{A}\prec\mathrm{B}},\Gamma^{\mathrm{B}\prec\mathrm{A}})$ and $H(Q)\leq 1$.
As a result, the previous approach does not allow us to show the equivalence of choice between $\widetilde{\mathcal{S}}$ and $\widetilde{\mathcal{S}}'$ in this situation, which we were indeed unable to prove.

Nonetheless, we stress that any possible differences in tightness between the entropic inequalities arising here from one particular probability structure or another are not due to stricter physical assumptions (i.e., the existence of joint probability distributions), but are rather due to different outer approximations of the true entropy cone (or polyhedron) via Shannon inequalities. 
We remark, however, that the choice of a minimal probability structure is computationally easier to handle due to the much lower number of variables; for example, compare the case $\mathcal{S}=\mathcal{M}$ in Eq.~\eqref{eq:shannonCone_counterfact}, where $|\mathcal{S}^{\rm c}| = 13$ and thus $\Gamma^{\mathcal{S}} \subset \mathbb{R}^{13}$ (and where the entropy vectors to be considered are effectively $12$-dimensional, since $H(\emptyset)$ is fixed to be $0$), with the corresponding case for $\mathcal{S}'=\{S\}$, where $\Gamma_S \subset \mathbb{R}^{2^8}=\mathbb{R}^{256}$ (with effectively $255$-dimensional entropy vectors).
For an extensive discussion of the role of such constraints in the computation of tighter approximations to the entropy cone we refer the reader to Ref.~\cite{BMC2016}.

\end{document}